\documentclass[a4paper,12pt]{article}
\pdfoutput=1
\usepackage{graphicx, rotating,slashed,ifpdf,amssymb,amsmath}
\usepackage[dvipsnames]{xcolor}
\usepackage{cancel}

\usepackage{charter}

\ifx\pdfoutput\undefined
\usepackage[dvips,bookmarks=false]{hyperref}	
\else
\usepackage{hyperref}	
\fi
\hypersetup{colorlinks,bookmarksopen,bookmarksnumbered,citecolor=verdes,
linkcolor=blus,pdfstartview=FitH,urlcolor=rossos}
\def\myurl#1#2{\href{http://#1}{#2}}
\def\hhref#1{\href{http://arxiv.org/abs/#1}{#1}} 

\usepackage{amsmath}
\usepackage{slashed}

\newcommand{\mb}[1]{\mbox{\normalsize\boldmath $#1$}}
\newcommand{\rhob}{{\mb{\rho}}}

\newcommand{\beq}{\begin{equation}}
\newcommand{\eeq}{\end{equation}}
\newcommand{\fig}[1]{~\ref{fig:#1}}
\newcommand{\be}{\begin{equation}}
\newcommand{\ee}{\end{equation}}

\newcommand{\nubarnu}{\raisebox{1ex}{\hbox{\tiny(}}\overline\nu\raisebox{1ex}{\hbox{\tiny)}}\hspace{-0.5ex}}

\newcount\Mac  \Mac=1  
\newcommand{\ifMac}[2]{\ifnum\Mac=1 #1 \else #2 \fi}
\def\putps(#1,#2)(#3,#4)#5#6{\ifnum\Mac=1 \put(#1,#2){\special{picture #5}}
\else  \put(#3,#4){\includegraphics{#6}} \fi}

\newcommand{\One}{\hbox{1\kern-.24em I}}

\newcommand{\pb}{\,{\rm pb}}
\newcommand{\GeV}{\,{\rm GeV}}

\newcommand{\MeV}{\,{\rm MeV}}
\newcommand{\eV}{\,{\rm eV}}

\newcommand{\eq}[1]{~{\rm(\ref{eq:#1})}}

\newcommand{\lascia}[1]{}
\makeatletter
\def\art{\@ifnextchar[{\eart}{\oart}}
\def\eart[#1]#2#3#4#5#6{{\rm #2}, {#3 #4} {\rm (#6) #5} [arXiv:{\hhref{#1}}]}
\def\hepart[#1]#2{{\rm #2, arXiv:\hhref{#1}}}
\newcommand{\oart}[5]{{\rm #1}, {#2 #3} {\rm (#5) #4}}

\newcounter{alphaequation}[equation]
\def\thealphaequation{\theequation\hbox to
0.6em{\hfil\alph{alphaequation}\hfil}}
\def\eqnsystem#1{
\def\@eqnnum{{\rm (\thealphaequation)}}
\def\@@eqncr{\let\@tempa\relax \ifcase\@eqcnt \def\@tempa{& & &} \or
  \def\@tempa{& &}\or \def\@tempa{&}\fi\@tempa
  \if@eqnsw\@eqnnum\refstepcounter{alphaequation}\fi
\global\@eqnswtrue\global\@eqcnt=0\cr}
\refstepcounter{equation} \let\@currentlabel\theequation \def\@tempb{#1}
\ifx\@tempb\empty\else\label{#1}\fi
\refstepcounter{alphaequation}
\let\@currentlabel\thealphaequation
\global\@eqnswtrue\global\@eqcnt=0 \tabskip\@centering\let\\=\@eqncr
$$\halign to \displaywidth\bgroup \@eqnsel\hskip\@centering
$\displaystyle\tabskip\z@{##}$&\global\@eqcnt\@ne
\hskip2\arraycolsep\hfil${##}$\hfil& \global\@eqcnt\tw@\hskip2\arraycolsep
$\displaystyle\tabskip\z@{##}$\hfil
\tabskip\@centering&\llap{##}\tabskip\z@\cr}

\def\endeqnsystem{\@@eqncr\egroup$$\global\@ignoretrue} \makeatother

\def\diag{\mathop{\rm diag}}

\def\circa#1{\,\raise.3ex\hbox{$#1$\kern-.75em\lower1ex\hbox{$\sim$}}\,}

\def\ss0{{\scriptscriptstyle 0}}
\def\1l{{\scriptscriptstyle (1)}}
\def\2l{{\scriptscriptstyle (2)}}

\usepackage{multicol}
\usepackage{color}
\definecolor{rosso}{cmyk}{0,1,1,0.4}
\definecolor{rossos}{cmyk}{0,1,1,1}
\definecolor{rossoc}{cmyk}{0,1,1,0.2}
\definecolor{blu}{cmyk}{1,1,0,0.3}
\definecolor{blus}{cmyk}{1,1,0,0.6}
\definecolor{bluc}{cmyk}{1,1,0,0.1}
\definecolor{verde}{cmyk}{0.92,0,0.59,0.25}
\definecolor{verdec}{cmyk}{0.92,0,0.59,0.15}
\definecolor{verdes}{cmyk}{0.92,0,0.59,0.4}
\definecolor{grigio}{cmyk}{0,0,0,0.07}
\definecolor{rosa}{cmyk}{0,0.1,0.1,0.02}
\definecolor{rosino}{cmyk}{0,0.05,0.05,0.02}
\definecolor{rosas}{cmyk}{0,0.3,0.25,0.05}
\definecolor{celeste}{cmyk}{0.1,0,0,0.02}
\definecolor{giallino}{cmyk}{0,0,0.4,0.02}
\definecolor{rosso}{cmyk}{0,1,1,0.4}
\definecolor{rossos}{cmyk}{0,1,1,0.55}
\definecolor{rossoc}{cmyk}{0,1,1,0.2}
\definecolor{blu}{cmyk}{1,1,0,0.3}
\definecolor{bluc}{cmyk}{1,1,0,0.1}
\definecolor{blucc}{cmyk}{0.7,0.5,0,0}
\definecolor{viola}{cmyk}{0,1,0,0.6}
\definecolor{viola2}{cmyk}{0,1,0.2,0.6}
\definecolor{verde}{cmyk}{0.92,0,0.59,0.25}
\definecolor{verdec}{cmyk}{0.92,0,0.59,0.15}
\definecolor{verdes}{cmyk}{0.92,0,0.59,0.4}
\definecolor{verdino}{cmyk}{0.12,0,0.09,0.05}
\definecolor{giallo}{cmyk}{0,0,1,0}
\definecolor{gialloverde}{cmyk}{0.44,0,0.74,0}

\font\tenrsfs=rsfs10 at 12pt

\font\sevenrsfs=rsfs7
\font\fiversfs=rsfs5
\newfam\rsfsfam
\textfont\rsfsfam=\tenrsfs
\scriptfont\rsfsfam=\sevenrsfs
\scriptscriptfont\rsfsfam=\fiversfs
\def\mathscr#1{{\fam\rsfsfam\relax#1}}

\def\beq{\begin{equation}}
\def\eeq{\end{equation}}
\def\bea{\begin{eqnarray}}
\def\eea{\end{eqnarray}}
\def\bac{\begin{array} {c}}
\def\ea{\end{array}}

\def\dm2{\delta m^2}
\def\dv2{\delta v^2}

\begin{document}
\begin{flushright} 
\footnotesize
SACLAY--T13/040
\end{flushright}

\color{black}
\vspace{1cm}
\begin{center}
{\LARGE\bf \color{Magenta}PPPC 4 DM$\nu$:}\\[3mm]
{\LARGE\bf \color{Magenta} A Poor Particle Physicist Cookbook}\\[3mm] 
{\LARGE\bf \color{Magenta} for Neutrinos from Dark Matter annihilations}\\[3mm] 
{\LARGE\bf \color{Magenta} in the Sun}\\
\bigskip\color{black}\vspace{0.6cm}{
{\large\bf  Pietro Baratella$^a$, Marco Cirelli$^b$, Andi Hektor$^{c,d}$,\\[1mm] Joosep Pata$^c$, Morten Piibeleht$^c$ {\rm and} Alessandro Strumia$^{c,e}$}
} \\[7mm]
{\it  (a) \href{http://www.sns.it}{Scuola Normale Superiore} and INFN, Piazza dei Cavalieri 7, 56126 Pisa, Italy}\\[1mm]
{\it  (b) \href{http://ipht.cea.fr/en/index.php}{Institut de Physique Th\'eorique}, CNRS URA 2306 \& CEA-Saclay, 
	91191 Gif-sur-Yvette, France}\\[1mm]
{\it (c) \href{http://www.kbfi.ee}{National Institute of Chemical Physics and Biophysics}, Ravala 10, Tallinn, Estonia}\\[1mm]
{\it (d) \href{http://www.hip.fi}{Helsinki Institute of Physics}, P.O. Box 64, FI-00014, Helsinki, Finland}\\[1mm]
{\it (e) \href{http://www.df.unipi.it}{Dipartimento di Fisica dell'Universit{\`a} di Pisa} and INFN, Italy}\\[1mm]
\end{center}
\bigskip
\bigskip
\bigskip
\vspace{1cm}

\centerline{\large\bf\color{blus} Abstract}

\begin{quote}
We provide ingredients and recipes for computing neutrino signals of TeV-scale Dark Matter (DM) annihilations in the Sun.
For each annihilation channel and DM mass we present the energy spectra of neutrinos at production, including:  state-of-the-art energy losses of primary particles in solar matter, secondary neutrinos, electroweak radiation. 
We then present the spectra after propagation to the Earth, including (vacuum and matter) flavor oscillations and interactions in solar matter. 
We also provide a numerical computation of the capture rate of DM particles in the Sun.
These results are \myurl{www.marcocirelli.net/PPPC4DMID.html}{available in numerical form}.
\end{quote}

\newpage

\tableofcontents

\section{Introduction}

The Dark Matter (DM) particles that constitute the halo of the Milky Way have a small but finite probability of scattering with a nucleus of a massive celestial body (e.g.\ the Sun or the Earth) if their orbit passes through it. If their velocity after the scattering is smaller than the escape velocity from that body, they become gravitationally bound and start orbiting around the body. Upon additional scatterings, they eventually sink into the center of the body and accumulate, building up a local DM overdensity concentrated in a relatively small volume. There they annihilate into  Standard Model particles, 
giving origin to fluxes of energetic neutrinos~\cite{DMnu}. These neutrinos are the only species that can emerge, after experiencing  oscillations and interactions in the dense matter of the astrophysical body. The detection of high-energy neutrinos from the center of the Sun or the Earth, on top of the much lower energy neutrino flux due to nuclear fusion or radioactive processes, would arguably constitute one of the best proverbial smoking guns for DM, as there are no known astrophysical processes  able to mimic it (except possibly for the flux of neutrinos produced in the atmosphere of the Sun, which however are expected to have a different spectral shape~\cite{corona}). 

\medskip

While the above idea is rather old, recently it has gained more interest since large neutrino telescopes ({\sc Antares}, {\sc Icecube} and its planned extension {\sc Pingu}, {\sc SuperKamiokande} and its proposed upgrade {\sc HyperKamiokande}, the future {\sc Km3Net}) are reaching the sensitivity necessary to probe one of the most interesting portions of the parameter space, in competition with other DM search strategies. 

Along the years there have been several computations of the ingredients necessary to predict the signal in a detector, in particular  the DM capture rate~\cite{capture} and  the spectra of neutrinos~\cite{earlyspectra,nuDM1,WIMPSIM,Rott:2012qb,Bernal:2012qh}. In~\cite{nuDM1} a subset of the authors of this paper had computed the neutrino spectra: their production in solar matter and their propagation subject to interactions and flavor oscillations. Similar results, but in an event-based approach, had been subsequently obtained by {\sc Wimpsim}~\cite{WIMPSIM}. 

\medskip

In this paper we improve the computation of the neutrino energy 
spectra with respect to previous computations and in particular to~\cite{nuDM1} 
in various ways: 
\begin{itemize}
\item we consider a full list of two-body SM annihilation channels; 
\item we implement a better description of the energy losses of the primary particles in solar matter, including 
secondary neutrinos produced by scattered solar matter;
\item we extend to low energy the neutrino spectra in order to include the large flux of neutrinos generated by 
stopped $\mu,\pi,K$ (an effect pointed out in~\cite{Rott:2012qb,Bernal:2012qh});
\item we take into account EW bremsstrahlung;  
\item we update the neutrino oscillation parameters adding the recently discovered $\theta_{13}\approx 8.8^\circ$
and the two possible mass hierarchies of neutrinos (normal or inverted).
\end{itemize}

\bigskip

\noindent Let us now review the main steps of the computation and present the results we provide. 
 The final energy spectrum of the neutrino flux at the detector location is written as
\begin{equation}
\frac{d\Phi_\nu}{dE_\nu} = \frac{\Gamma_{\rm ann}}{4\pi d^2} \, \frac{dN_\nu}{dE_\nu}
\label{eq:nuflux}
\end{equation}
where $d$ is the  Sun--Earth distance, 
$\Gamma_{\rm ann}$ is the total DM annihilation rate in the Sun,
$dN_\nu/dE_\nu$ is the energy spectrum of $\nu_{e,\mu,\tau}$ and $\bar\nu_{e,\mu,\tau}$
produced per DM annihilation after taking into account all effects.
The computation proceeds as follows:
\begin{enumerate}

\item 
The total rate of DM annihilations $\Gamma_{\rm ann}$ in the Sun is computed in section~\ref{Gammaann} in terms
of the spin-independent and spin-dependent DM cross sections with nucleons.

\item 
In section~\ref{dNnudE}
we compute the energy spectra of $\nu$ and $\bar\nu$ at production for DM annihilations into
\beq \label{eq:channels}
\begin{array}{ll}
e_L^+e_L^-,\ 
e_R^+e_R^-,\
\mu_L^+\mu_L^-,\ 
\mu_R^+\mu_R^-,\ 
\tau_L^+\tau_L^-,\ 
\tau_R^+\tau_R^-,
\nu_e\bar\nu_e,\nu_\mu\bar\nu_\mu, \nu_\tau\bar\nu_\tau,
\\[3mm]
q \bar q, \ 
c \bar c, \ 
b \bar b, \  
t \bar t, \ 
\gamma \gamma,\ 
g g,  
W_L^+ W_L^-,\ 
W_T^+W_T^-,\ 
Z_LZ_L,\ 
Z_T Z_T,  hh
\end{array}
\eeq
where  $q ={u,d,s}$ denotes a light quark, 
the subscripts $L,R$ on leptons denote $L$eft or $R$ight-handed polarizations,
the subscript $L,T$ on massive vectors denote $L$ongitudinal or $T$ransverse polarization. $h$ denote the higgs boson, for which we assume a mass of 125 GeV.

Decays, hadronization, energy losses in matter and secondary neutrinos
generated by matter particles scattered by DM decay products (an effect ignored in previous computations)
are computed running {\sc Geant4}~\cite{Geant} using the EU Baltic Grid facilities. 
As a check, the first three effects are also independently computed with {\sc Pythia}~\cite{Pythia8}, modified by us to include 
energy losses in matter.

\item As described in section~\ref{sec:EWDM},
we correct the spectra at production including the dominant 
electroweak radiation effects (not included in MonteCarlo codes) enhanced by logarithms of $M_{\rm DM}/M_W$,
as described in~\cite{EWDM}.  EW corrections depend on the polarisation of the SM particles,
which is why we need to specify it in eq.\eq{channels}.

\item As described in section~\ref{propagation},
we propagate the fluxes of $\nu$ and $\bar\nu$ at production, around the center of the Sun, to the Earth taking into account
oscillations, matter effects, absorption and regeneration from collisions with matter using
the neutrino density matrix formalism as in~\cite{nuDM1}.

\end{enumerate}

\noindent The final results are presented and briefly discussed in section~\ref{sec:finalresult}. In section~\ref{sec:concl} we conclude.

\medskip

\noindent  The main numerical outputs of the computation are given on the \myurl{www.marcocirelli.net/PPPC4DMID.html}{PPPC 4 DM ID website} (`DM$\nu$' section).


\section{The DM annihilation rate}
\label{Gammaann}

In this section we compute the DM solar annihilation rate, which sets the overall normalization of the expected neutrino flux. We essentially revisit the calculations in~\cite{capture}. For definiteness, we focus on the case of the Sun, although we mention in passing the modifications needed for the Earth.

\medskip

The number $N$ of DM particles accumulated inside the Sun varies with time under the action of different competing processes: 1) DM particles are captured from the halo ($N$ is thus increased by one unit) via the multiple scattering processes discussed above, 2) DM can pair annihilate (hence $N$ decreases by two units) and 3) DM particles can be ejected ($N$ decreases by one unit) by a hard scattering on a hot nucleus of the interior of the body, i.e. the inverse process with respect to capture. In formul\ae:
\beq 
\frac{dN}{dt} = \Gamma_{\rm capt} -2\Gamma_{\rm ann} - \Gamma_{\rm evap}
\label{eq:DMcaptured}
\eeq
where $\Gamma_{\rm capt}$ is the capture rate, $\Gamma_{\rm ann}$ is the DM annihilation rate and $\Gamma_{\rm evap}$ is the DM evaporation (or ejection) rate.

The evaporation process is important only for DM lighter than a few GeV~\footnote{This can be easily understood~\cite{capture}: if one requires that the speed of a DM particle thermalized with the interior of the Sun $v_{\rm DM} \sim \sqrt{2 T_\odot/M_{\rm DM}}$ be smaller than the escape velocity from the center of the Sun $v_{\odot\rm esc} \simeq 1387$ km/s, one obtains the condition $M_{\rm DM} \gtrsim 0.15$ GeV. A more realistic computation takes into account that DM particles actually follow a (Maxwellian) velocity distribution and compares the time it takes to deplete the high-velocity tail (above the escape velocity) of such distribution with the age of the Sun, finding $M_{\rm DM} \gtrsim 5$ GeV. Yet more refined calculations find typically $M_{\rm DM} \gtrsim 4$ GeV~\cite{Busoni:2013kaa}.}, so that we will neglect it for all practical purposes in the following.

The annihilation rate is proportional to $N^2$: two DM particles annihilate (hence the square).
 It is given by
\beq \label{eq:Gammann}
\Gamma_{\rm ann} =  \frac12
\int d^3x \, n^2(\vec x)\, \langle \sigma v \rangle =
\frac{1}{2} C_{\rm ann} N^2,  
\eeq
where  $\langle \sigma v \rangle$ is the usual annihilation cross section averaged over the initial state\footnote{If DM is a real particle (e.g.\ a Majorana fermion) this is the usual definition of $\sigma$ and the factor $1/2$ takes into account the symmetry of the initial state.
If DM is a complex particle (e.g.\ a Dirac fermion) then $n \equiv n_{\rm DM} + n_{\overline{\rm DM}}$
(here assumed to be equal) and
the average over initial states is
$\sigma \equiv \frac{1}{4}(2\sigma_{{\rm DM}\,\overline{\rm DM}} + 
\sigma_{\rm DM~DM} + \sigma_{\overline{\rm DM}\overline{\rm DM}})$.
In many models, only ${\rm DM}\,\overline{\rm DM}$ annihilations are present, so that
$\sigma =  \sigma_{{\rm DM}\,\overline{\rm DM}}/2$.}
and
$n(\vec x)$ is the number density of DM particles at position $\vec x$ inside the Sun, such that the total number of DM particles is
 $N=\int d^3x \, n(\vec x)$.
After capture, subsequent scatterings thermalize the DM particles to the solar temperature $T_\odot$, such that
their density $n(\vec x)$ acquires the spherically symmetrical Boltzmann form
\beq n(r) = n_0 \, \exp[-M_{\rm DM}\, \phi(r)/T_\odot] \label{WIMPdistrib}\eeq
where $n_0$ is the central DM number density and 
$\phi(r) = \int_0^r dr\, G_N M(r)/r^2 $ is the Newtonian gravitational potential inside the Sun,
written in terms of the solar mass $M(r)$ enclosed within a sphere of radius $r$.
Taking for simplicity the matter density in this volume to be constant and equal to the central density $\rho_\odot$, all the integrals can be explicitly evaluated . One finds that
DM particles are concentrated around the center of Sun,
\beq\label{eq:n(r)}
n(r) = n_0 \, e^{-r^2/r_{\rm DM}^2}, \quad {\rm with} \qquad  r_{\rm DM} = \left(\frac{3 T_\odot}{2 \pi \, G_N \rho_\odot \, M_{\rm DM}} \right)^{1/2}\approx
0.01 \, R_\odot \sqrt{\frac{100 \, {\rm GeV}}{M_{\rm DM}}}.
\eeq
Within this approximation, one obtains from eq.\eq{Gammann} 
\beq
C_{\rm ann}=
\langle\sigma v\rangle \left(\frac{G_N \, M_{\rm DM} \,\rho_\odot}{3\,T_\odot}\right)^{3/2}.
\eeq
Here $\rho_\odot = 151$ g/cm$^3$ and $T_\odot = 15.5 \ 10^6$ K are the density and the temperature of matter around the center of the Sun. The same expression would hold for other astrophysical bodies, adapting these two quantities.

\medskip

Neglecting $\Gamma_{\rm evap}$ and solving eq.\eq{DMcaptured} with respect to time one finds
\beq 
\label{eq:equilibrium}
\Gamma_{\rm ann} = \frac{\Gamma_{\rm capt}}{2} \tanh^2 \left(\frac{t}{\tau}\right)
\stackrel{t\gg\tau}{\simeq} \frac{\Gamma_{\rm capt}}{2}
\eeq
where  $\tau = 1/\sqrt{\Gamma_{\rm capt} C_{\rm ann}}$ is a time-scale set by the competing processes of capture and annihilation. At late times $t\gg \tau$ one can  approximate $\tanh(t/\tau)=1$. 
In the case of the Sun, the age of the body ($\sim$4.5 Gyr) and the typical values of the parameters in $\tau$ indeed satisfy this condition (in the case of the Earth this is not generally the case). Therefore one attains the last equality of eq.~(\ref{eq:equilibrium}).  
Physically, this means that the fast (compared to the age of the Sun) processes of capture and annihilation come to an equilibrium: any additional captured particle thermalizes and eventually is annihilated away.

\begin{figure}[t]
\begin{center}
$$\includegraphics[width=0.45\textwidth]{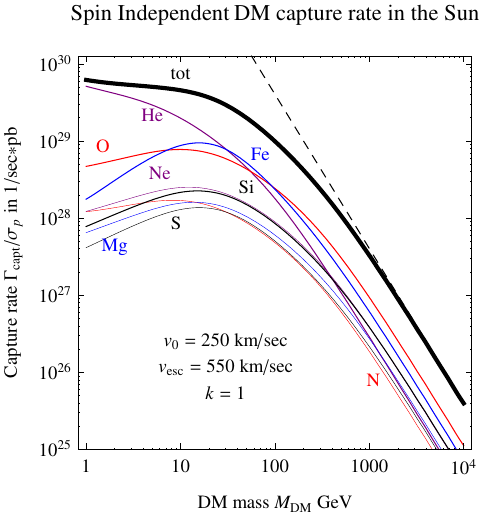}\qquad
\includegraphics[width=0.45\textwidth]{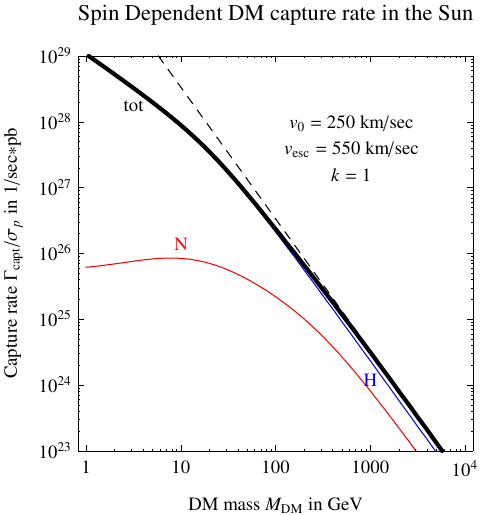}$$
\caption{\em {\bfseries DM capture rate in the Sun} $\Gamma_{\rm capt}$  of DM particles with mass $M_{\rm DM}$  assuming a
Spin-Independent (left) or a Spin-Dependent (right) DM cross section on matter.
We normalize assuming a cross section on protons $\sigma_p = 1\,{\rm pb}$ and we adopt the indicated values for the parameters of the galactic DM velocity distribution.
We show the total as well as the contributions from the most relevant individual elements in the Sun.
The dashed line is the analytical approximation valid when the DM is much heavier than the nuclei.
\label{fig:capturerate}}
\end{center}
\end{figure}

\subsection{The DM capture rate}

As a consequence of the attained equilibrium, the computation of the annihilation rate $\Gamma_{\rm ann}$ crucially depends on the computation of the capture rate $\Gamma_{\rm capt}$, which acts as a bottleneck. The computation of the latter proceeds by summing the contribution to capture of all the individual shells of matter located at position $r$ within the massive body. The  result is~\cite{capture}
\beq 
\Gamma_{\rm capt} = 
\frac{\rho_{\rm DM}}{M_{\rm DM}}\sum_i \sigma_i
\int_0^{R_\odot} dr~4\pi r^2~ n_i(r)
 \int_0^\infty dv~ 4\pi v^2 f_\odot(v) \frac{v^2 + v_{\odot\rm esc}^2}{v} \wp_i (v,v_{\odot\rm esc}).
 \eeq
Its derivation is lengthy: we will only sketch it in the following by illustrating the individual pieces of this equation. 
\begin{itemize}
\item $\rho_{\rm DM}/M_{\rm DM}$ corresponds just to the local number density of DM particles at the location of the capturing body.
\item The sum runs over all kinds of nuclei $i$ with mass $m_i$ and number density $n_i(r)$,
to be integrated over the volume of the Sun. We take the standard solar elemental abundances from~\cite{solarcomposition}. 
The factor $\sigma_i $ is the low-energy DM cross section on nucleus $i$, assumed to be isotropic.
In terms of the standard Spin Independent and Spin Dependent cross sections normalized on a single nucleon at zero momentum transfer\footnote{We assume that they
are equal for protons and neutrons and that they do not depend on the momentum transfer nor on the DM velocity.} (denoted $\sigma_{\rm SI}$ and $\sigma_{\rm SD}$)
we take $\sigma_{\rm H} = \sigma_{\rm SI} + \sigma_{\rm SD}$ for Hydrogen and
$\sigma_{\rm He} \simeq 4^4 \,\sigma_{\rm SI}  (m_p + M_{\rm DM})^2/(m_{\rm He} + M_{\rm DM})^2$ for 
the spin-less Helium. Analogously, for heavier nuclei of mass number $A$ one can take $\sigma_i \simeq A^2 \,  \sigma_{\rm SI} \, (m_A/m_p)^2\ (m_p+M_{\rm DM})^2/(m_A+M_{\rm DM})^2$ (spin independent contribution) + $\sigma_i \simeq A^2 \, \sigma_{\rm SD} \, (m_p+M_{\rm DM})^2/(m_A+M_{\rm DM})^2$ (spin dependent contribution). This latter SD formula should be complemented by weighting factors due to the spin content in the different nuclei. In practice, only Hydrogen is relevant for our case, so that a more careful evaluation is not necessary.

\item $f_\odot(v)$ is the angular average of the DM velocity distribution with respect to the {\em solar} rest frame neglecting the gravitational attraction of the Sun.
This quantity is connected to the $f(v)$ with respect to the {\em galactic} rest frame as
\begin{equation}
\label{eq:fsunav}
f_\odot(v) =  \frac{1}{2} \int_{-1}^{+1} dc  ~f(\sqrt{v^2 + v_\odot^2 + 2 v v_\odot c})
\end{equation}
in terms of the solar velocity $v_\odot\approx 233\,{\rm km/s}$ (here $c$ is the cosine of the angle between $v$ and $v_\odot$). 
Observations and numerical simulations suggest that $f(v)$ can be parameterized as~\cite{fv}
\beq
 f(v)= N \left[\exp\left(\frac{v_{\rm esc}^2-v^2}{kv_0^2}\right)-1\right]^k\theta(v_{\rm esc}-v)\,
 \eeq
 with normalisation $\int_0^\infty dv~4\pi v^2 f(v)=1$ and parameters
\beq 220\,{\rm km/s}<v_0<270\,{\rm  km/s},\qquad
450\,{\rm km/s}< v_{\rm esc}< 650\,{\rm  km/s},\qquad
1.5<k<3.5.\eeq
Here $v_{\rm esc}$ is the escape velocity from the Galaxy (not to be confused with $v_{\odot{\rm esc}}$, the escape velocity from the Sun) at the location of the solar system.  
For $k=0$, $f(v) $ reduces 
to a Maxwell-Boltzmann distribution with a sharp cut-off at $v<v_{\rm esc}$:
$f( v) = N\times  e^{-v^2/v_0^2}\theta(v_{\rm esc}-v)$.
For $k>0$ the cut-off gets smoothed.
The resulting $f_\odot (v)$
are normalised such that
$\int_0^\infty dv~4\pi v^2 f_\odot(v)=1$. The actual choice of the DM velocity distribution turns out to have a rather limited impact, see e.g.~\cite{Choi:2013eda} for a recent analysis.

\item The gravity of the Sun is taken into account by
$v_{\odot\rm esc}(r)$, which is the escape velocity (from the Sun) at radius $r$, such that $v^2 + v_{\odot\rm esc}^2$ is the squared velocity that a DM particle acquires when
arriving at $r$ from a very distant point.
One here neglects the effect of other bodies in the solar system, which presumably is a good approximation~\cite{DMnuplanets}.

\item The probability that a DM particle, with velocity $v$ far from the Sun,
gets captured when scattering on a nucleus located
where the escape velocity is $v_{\odot\rm esc}$ is given in first approximation by 
\beq
\wp_i (v,v_{\odot\rm esc}) =\max\bigg(0, \frac{\Delta_{\rm max} - \Delta_{\rm min}}{\Delta_{\rm max}}\bigg)
\label{eq:probcapt}\eeq
where $\Delta_{\rm max}=4 m_i M_{\rm DM}/(M_{\rm DM}+m_i)^2$ and  
$\Delta_{\rm min}=v^2/(v^2+v_{\odot\rm esc}^2)$ are the maximal and minimal fractional energy loss $\Delta E/E$ that a particle can suffer in the scattering process, provided that it is captured. 
I.e., the probability is computed as the ratio of the size of the interval in energy losses leading to capture ($\Delta E_{\rm min} <\Delta E< \Delta E_{\rm max}$) relative to the whole possible interval ($0 <\Delta E<\Delta E_{\rm max}$), assuming  a flat distribution of the scattering cross section in energy. 
In general, however, one needs to introduce the form factors $F_i(\Delta E)$ that take into account the nuclear response as function of the
momentum transfer. Explicitly, $|F_i(\Delta E)|^2=e^{-\Delta E/E_0}$, with $E_0=5/2m_i r_i^2$ for spin-independent and $E_0 = 3/2m_i r_i^2$ for spin-dependent scattering (here $r_i \sim \sqrt{3/5}\, A_i^{1/3} \, 1.23\ {\rm fm} \simeq 0.754\, \cdot 10^{-13}\, {\rm cm} \, (m_i/{\rm GeV})^{1/3}$ is the effective radius of a nucleus with mass number $A_i$ and mass $m_i$).
The numerator of the  `ratio of sizes' becomes then an integral of the form factor over the energy loss $\Delta E$:
\beq
\wp_i (v,v_{\odot\rm esc}) = \frac{1}{E \, \Delta_{\rm max}} \int_{E\, \Delta_{\rm min}}^{E\, \Delta_{\rm max}} {\rm d}(\Delta E)\, |F_i(\Delta E)|^2,
\label{eq:probcaptrefined}\eeq

Eq.~(\ref{eq:probcapt}) and\eq{probcaptrefined} 
mean that the fraction of scatterings that lead to capture is largest for nuclei with mass $m_i$ comparable to the DM mass $M_{\rm DM}$ ($\Delta_{\rm max}$ is maximized) and for DM particles that are slow (small $v$) and in the central regions of the body (large $v_{\odot \rm esc}$).

\end{itemize}

\bigskip

Fig.\fig{capturerate}a shows the capture rate in the Sun having assumed a Spin Indipendent cross section 
$\sigma^{\rm SI}_p = 1 \pb$ on protons. 
One sees that several elements contribute 
Fig.\fig{capturerate}b shows the Spin Dependent capture rate, with the corresponding assumption $\sigma^{\rm SD}_p = 1 \pb$ on protons. 
Only Hydrogen matters for this kind of capture, with a very small contribution from Nitrogen.

\medskip

The dotted lines in fig.\fig{capturerate} are simple approximations valid in the limit of heavy DM, $M_{\rm DM} \gg m_i$.
In such a limit  DM can be captured only if it is very slow, 
\beq v\stackrel{M_{\rm DM}\gg m_i}{<} 2v_{\odot\rm esc}\sqrt{m_i/M_{\rm DM}} .\eeq 
Thereby the capture rate is proportional to $1/M_{\rm DM}^2$ and can be approximated as 
\beq \Gamma_{\rm capt} \stackrel{M_{\rm DM}\gg m_i}{\simeq}  \frac{\rho_{\rm DM}}{M_{\rm DM}^2 } 
4 \pi f_\odot(0)
\sum_i m_i \sigma_i  I_i \label{eq:GammaCaptApprox}
\eeq
where
\beq 
	{I}_i=\int_0^{R_\odot} 4\pi r^2 n_i(r) \left[ \frac{1}{2}\left(\frac{E_{0i}}{m_i}\right)^2-\frac{E_{0i}}{m_i}e^{-2 m_i v_{\odot\rm esc}^2(r)/E_{0i}} \left( \frac{E_{0i}}{2m_i}+v_{\odot\rm esc}^2(r) \right) \right] dr
\eeq
In the limit of negligible form factors, $E_{0i} \gg m_i$, the term in square brackets simplifies to $v_{\odot\rm esc}^4$.

The integrals ${I}_i$ are adimensional in natural units, and their values are given in table~\ref{tab:asymptInteg}
for the main capturing elements.
Inserting their values we find  
\beq 
\label{eq:GammaCaptApproxNumbers}
\Gamma_{\rm capt} \simeq
\frac{ 5.90 \cdot 10^{26}}{\rm sec} \left(\frac{\rho_{\rm DM}}{0.3 \, \frac{\rm GeV}{\rm cm^3}}  \right) 
\left(\frac{100\, {\rm GeV}}{M_{\rm DM}} \right)^2 
\left(\frac{270 \, \frac{\rm km}{\rm sec}}{ v^{\rm eff}_0} \right)^3 
\frac{\sigma_{\rm SD} + 1200\ \sigma_{\rm SI} }{{\rm pb}}.
\eeq
We here parameterised $f_\odot(0) = \left(1/\pi^{1/2} v_0^{\rm eff}\right)^3$, 
such that the parameter $v^{\rm eff}_0$ coincides with the parameter $v_0$ for a
Maxwellian distribution with no cut-off and neglecting solar motion.

\begin{table}[t]
\begin{center}
\begin{tabular}{ l l l l l }
\hline\hline
SD capture & H & $1.60 \cdot 10^{47}$ & N & $2.42 \cdot 10^{43}$ \\
\hline
SI capture & He & $2.00 \cdot 10^{46}$ & N & $2.84 \cdot 10^{43}$ \\
& O & $7.34 \cdot 10^{43}$ & Ne & $1.03 \cdot 10^{43}$ \\
& Mg & $3.01 \cdot 10^{42}$ & Si & $1.98 \cdot 10^{42}$ \\
& S & $5.73 \cdot 10^{41}$ & Fe & $8.87 \cdot 10^{40}$ \\
\end{tabular}
\end{center}
\caption{\em \label{tab:asymptInteg} Value in natural units of $I_i$, for the main elements of the Sun.}
\end{table}

\bigskip

\noindent We also provide 
a numerical function which gives the annihilation rate $\Gamma_{\rm ann} = \Gamma_{\rm capt}/2$ in terms of $M_{\rm DM}$, the standard DM cross sections on a single nucleon $\sigma_{\rm SD}$ and $\sigma_{\rm SI}$ (as defined above) and the parameters of the velocity distribution $v_0$, $v_{\rm esc}$, $k$.

\section{The neutrino energy spectra at production}\label{dNnudE}

In this section we discuss the computation of the neutrino fluxes emerging from the DM annihilation process. Neutrinos are produced at several stages of the hadronic and leptonic cascades originating from the primary  particles produced
by annihilations, and these cascades develop within the dense matter of the astronomical body (the Sun). Such an environment has important consequences in determining the spectra. 
 For example, some byproducts of the cascade can be absorbed by the surrounding matter; some others will be just `slowed down'; others are negligibly affected. It is therefore important to model in the best possible way all these effects. We adopted and compared two strategies: 

\begin{itemize}
\item[1)]  As described in section~\ref{Pythia},
we modified the public MonteCarlo code {\sc Pythia}  to include the effects of cascading energy losses with matter and absorption.
The disadvantage is that in this way we cannot include the neutrinos emitted by matter particles scattered by products of DM annihilation.

\item[2)] As described in section~\ref{Geant}, we run the  {\sc Geant4} code, dedicated to particle interactions with matter.
The disadvantage is that {\sc Geant4}  is much more time-consuming than {\sc Pythia}:
to reach a reasonable statistics we employ the computational resources of the Baltic grid.
\end{itemize}
In section~\ref{sec:MCcomparison} we compare the two results, interpret them and decide to adopt the {\sc Geant4} result. 
In section~\ref{sec:EWDM} we describe how we subsequently add electroweak bremsstrahlung corrections,
which are not included in any Monte Carlo code.

\subsection{Neutrino spectra with MonteCarlo codes}

\subsubsection{{\sc Pythia}}\label{Pythia}
We here describe how we modified the {\sc Pythia} public code to include the effect of cascading in matter
in an event-by-event basis.
We improved with respect to our previous computation~\cite{nuDM1} where we had only included averaged energy losses in matter. Moreover, we here use {\sc Pythia} version 8.176 (version 6.2 was used in~\cite{nuDM1}).

The process of particles with mass $m$ and initial energy $E_0 \gg m$ decaying with life-time $\tau$, while at the same time losing energy due to interactions with the surrounding matter, is described by a differential equation for their energy $E$ and their number $n$: 
\beq \frac{dE}{dt}  =- \alpha - \beta E
,\qquad
\frac{dn}{dt}  = - \frac{n}{\tau}\frac{m}{E};\eeq
that is: the energy loss of particles in matter can by approximated
as a constant term $\alpha$ 
plus a term $\beta$ proportional to their energy, while their number $n$ is just governed by $\tau$ (the factor $m/E$ takes into account time dilatation).
Therefore
\beq \frac{dn}{dE}=\frac{dn/dt}{dE/dt}= \frac{n m}{\tau E(\alpha + \beta E)}\eeq
is solved by
\beq \label{eq:stopart}
n(E) = n(E_0) \left(\frac{\beta + \alpha/E_0}{\beta + \alpha/E}\right)^{m/\tau\alpha}.\eeq
We thereby instruct {\sc Pythia} to perform the decay of particles  by first replacing their initial energy $E_0$
with a lower energy $E$ randomly chosen according to the above distribution
produced by energy losses, eq.\eq{stopart}, and then letting them decay.
This equation takes different specific forms according to the particle considered, as we discuss now.

\subsubsection*{Dominant $\beta$: stopping of hadrons}
The stopping of hadrons is well approximated by $\alpha=0$.
In such a case eq.\eq{stopart} simplifies to
\beq
E(t) = E_0 e^{-\beta t},\qquad
\frac{n(E)}{n(E_0)} = \exp\bigg[\frac{E_{\rm cr}}{E_0} - \frac{E_{\rm cr}}{E(t)} \bigg]  = e^{x_0-x}
\eeq
where $E_{\rm cr}= m /\beta\tau$ is a critical energy below which energy losses are negligible.
The distribution of energies at decay is exponential in
$x= E_{\rm cr}/E$  (with $x_0\equiv E_{\rm cr}/E_0$).  
In the center of the Sun one has $E_{\rm cr}\approx 250\GeV$ for charmed hadrons and
$E_{\rm cr} \approx 470\GeV$ for $b$-hadrons~\cite{earlyspectra}.
This improves over our previous computation~\cite{nuDM1} where we used a constant
 average energy at decay, given by
\beq \langle E\rangle = \int E ~dn
=E_{\rm cr} \int n \frac{dE}{E} = E_{\rm cr} \int_{x_0}^\infty e^{x_0-x}\frac{dx}{x}.
\eeq
Neutrons and negatively charged pions need however a peculiar treatment: once stopped, instead of decaying, they are absorbed by matter.
Neutrons are  absorbed by protons, forming deuteron.
The $\pi^-$ are quickly Coulomb-captured by heavy nuclei in the dense and hot solar core environment, forming pionium which typicallys decay into a neutron and  photons~\cite{Ponomarev:1973ya}.
Since no neutrinos are produced in final states, to implement this phenomenology we simply modified our {\sc Pythia} code in such a way that neutrons and pions are removed and not decayed.

\subsubsection*{Dominant $\alpha$: stopping of charged leptons}
The energy loss of a charged lepton can often be approximated as an energy-independent term (i.e. taking $\beta = 0$), which in the center of the Sun equals to
\beq \frac{dE}{dt}=-\alpha \approx - \frac{0.8~10^{10}\GeV}{\sec}.\eeq
The general solution of eq.\eq{stopart} simplifies then to
\beq E(t) = E_0 - \alpha   t,\qquad   n(E)  =n(E_0) \left(\frac{E}{E_0}\right)^{m/\tau\alpha}.\eeq
The exponent $m/\tau\alpha$ equals $1/176000$ for muons (which therefore lose almost all their energy before decaying) and equals $660$ for taus (which therefore decay before losing a significant fraction of their initial energy).
The average energy at decay employed in our previous computation is
\beq \langle E\rangle = \int E~dn=  \frac{E_0}{1+\tau\alpha/m}.
\eeq

\subsubsection{\sc Geant}
\label{Geant}
{\sc Geant4} is a toolkit for the simulation of the passage of particles through matter~\cite{Geant}, a very suitable tool for modeling cascades in the environment of the solar core. We follow the approach of \cite{Rott:2012qb,Bernal:2012qh}.
Hadronization of the quarks and gluons produced by DM annihilations is performed by {\sc Pythia};
the stable and metastable hadronization products ($e$, $\nu_{e,\mu,\tau}$, $\mu$, $K_{L0}$, $\pi$, $K$, $n$, $p$, $N_{\rm D, He}$) are injected into {\sc Geant4} which adds the effect of particle/matter interactions.
We model the matter around the solar core following the Solar Standard Model as discussed in Sec.~\ref{Gammaann}, but with a simplified chemical composition (we just include 74.7 \% by mass of H and 25.3 \% of He).
In our {\sc Geant4} code, we consider a sphere with radius of 1 km, that is big enough for our purposes: we verified that only neutrinos from the cascades can reach its surface, while 
secondary products are contained. The passage of particles through matter is simulated using the QGSP\_BERT physics list, see~\cite{Geant} for further details. Notice that neutrinos have no matter interactions nor oscillations in the QGSP\_BERT physics model of {\sc Geant4} (we will indeed discuss in detail how to deal with these effects in the subsequent neutrino propagation in matter in Sec.~\ref{propagation}).

\subsubsection{Results and comparisons}\label{sec:MCcomparison}

\begin{figure}[t]
\begin{center}
\includegraphics[width=0.31\textwidth]{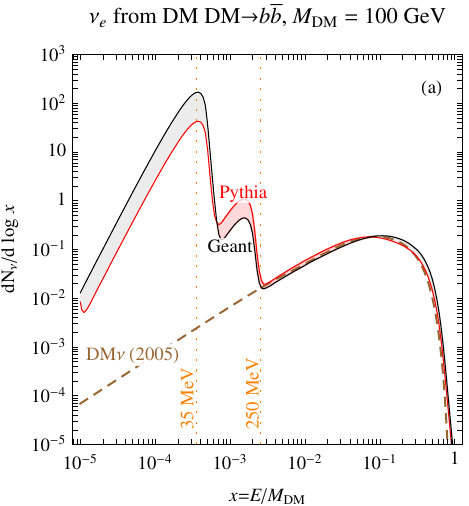} \quad
\includegraphics[width=0.31\textwidth]{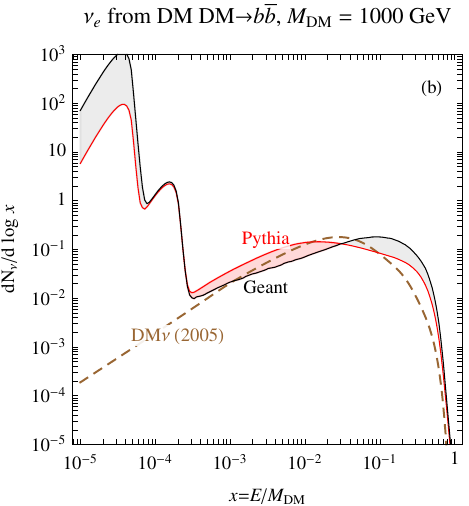} \quad
\includegraphics[width=0.31\textwidth]{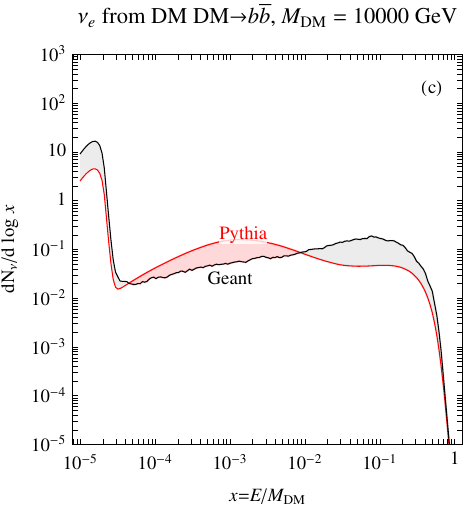}\\[5mm]
\includegraphics[width=0.31\textwidth]{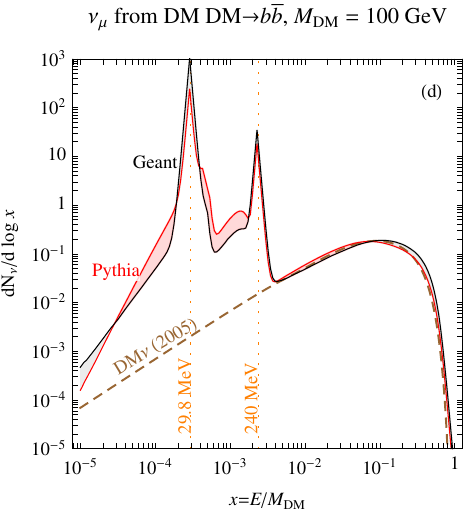} \quad
\includegraphics[width=0.31\textwidth]{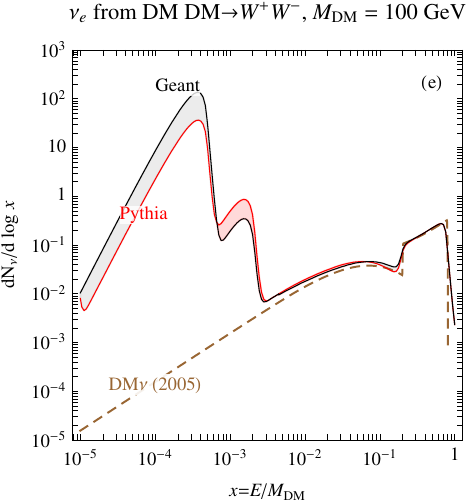} \quad
\includegraphics[width=0.31\textwidth]{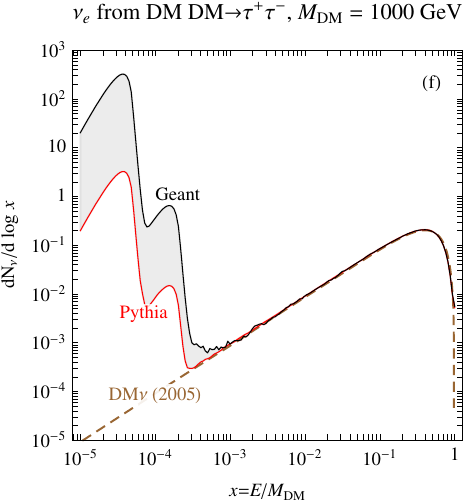}
\caption{\em {\bfseries Comparison} between the spectra obtained  {\bfseries with {\sc Pythia} and with {\sc Geant}}, for a few representative cases. See text for comments.
\label{fig:MCcomparison}}
\end{center}
\end{figure}

Fig.\fig{MCcomparison} presents a few examples of neutrino spectra obtained with the two MonteCarlo codes. It also shows (when available) the spectra previously computed in~\cite{nuDM1}.

In very general terms, moving from low to high $x = E/M_{\rm DM}$, the new spectra are characterized by some very pronounced low energy humps or spikes (missing in~\cite{nuDM1}), an intermediate energy smooth shoulder and, for some channels, a high energy peak. These features are easily understood. 
\begin{itemize}
\item The high energy peak occurs when DM annihilate into particles that directly decay into neutrinos. This is  visible in fig.\fig{MCcomparison} as a slanted mountain top in the $W^+W^-$ channel (panel (e)). A similar peak is of course present for the $ZZ$ channel (not shown). For large DM masses, i.e. for large boosts of the primary $W$ and $Z$, this feature smears into a smooth spectrum (see e.g. panel (a) of fig.~\fig{EWcomparison}). 

\item The smooth component of the spectrum arises from the neutrinos produced in the cascading event by primary and secondary particles (hadrons and leptons), that lose energy and rapidly decay. 
\item The low energy humps and spikes essentially arise from relatively long-lived particles that have been stopped in solar matter and then decay, essentially at rest.
More precisely, 
recalling that $n$, $\pi^-$, $\mu^-$ and $K^-$ are mostly absorbed or captured by matter, the low energy neutrino peaks arise from the following processes:
\begin{itemize}

\item $\pi^+ \to \mu^+ \nu_\mu$ decays, which produce
a monochromatic line in the $\nu_\mu$ spectra, at $E_\nu=29.8\MeV$  (visible in panel (d) of fig.\fig{MCcomparison}).
For numerical reasons we artificially broaden it.

\item $\mu^+\to \bar\nu_\mu \nu_e e^+$ decays, which
contribute to the $\nu_e,\bar\nu_\mu$ energy spectra producing
the typical three body decay bump with an end-point at $m_\mu/2\approx 53\MeV$ and
peaked at $E \approx m_\mu/3\approx  35 \MeV$ (visible in panel (a)).

\item $\mu^-\to \nu_\mu \bar\nu_e e^- $ decays, which similarly contribute to the $\bar\nu_e,\nu_\mu$ energy spectra,
although the resulting bump is about two orders of magnitude less intense
than the bump in $\nu_e,\bar\nu_\mu$. 
Indeed $\pi^-$ are absorbed by matter before decaying into $\mu^-$.

\item $K^+$ decays with $63 \%$ branching fraction into a monochromatic $\nu_\mu$ at $E_\nu=240\MeV$ (see panel (d)).
Three body decays have smaller branching ratios ($5.1\%$ BR into $\pi^0~e^+~\nu_e$ and $3.4\%$ BR into $\pi^0~\mu^+~\nu_\mu$)
producing bumps below about $m_K/2\approx 250\MeV$ (see e.g. panel (a)).

\item $K^-$ get absorbed, synthesised by nuclei into a $\Lambda$ which decays into nucleons and pions.
Their rare free decays negligibly affect the neutrino spectra.

\item 
Decays of $K^0_L$ into neutrinos are blocked by matter
effects that break the quantum coherence between $K^0_S$ and $K^0_L$,
such that $K^0_S$ are continuously regenerated and fastly decay hadronically.

\end{itemize}
\end{itemize}

While these general features are present both in the {\sc Pythia} and the {\sc Geant} spectra, the results of the two MonteCarlos differ  in the details. 
\begin{itemize}
\item For hadronic channels (e.g.\ $b \bar b$), {\sc Pythia} systematically produces a softer high-energy spectrum than {\sc Geant}, as clearly visible in panels (a), (b) and (c) (and (d) too). The difference is small for small DM masses and increases for large DM masses. For the leptonic channels, on the other hand, the two results agree (see e.g. panel (f)). 
This can probably be ascribed to a difference in the way {\sc Pythia} treats the energy losses of the energetic hadrons in the cascade.

\item {\sc Pythia} systematically underestimates the low energy humps and spikes, as it is clearly visible for instance in panel (f).\footnote{Panels (a) and (e) actually show a small difference in the opposite direction, of which we cannot trace the origin. We tentatively attribute it to small calibration discrepancies among the codes. It is present only at relatively low DM masses.} This is expected since, as we anticipated, {\sc Pythia} includes only the neutrinos produced by the decays of pions, kaons and muons in the DM-originated shower. {\sc Geant}, instead, follows the fate of all the particles produced in the scattered matter, including therefore additional pions, kaons and muons that then decay into neutrinos. \\

\end{itemize}

\noindent Based on the considerations above, we adopt the {\sc Geant} spectra.

\subsection{Adding electroweak bremsstrahlung}\label{sec:EWDM}

\begin{figure}[t]
\begin{center}
\includegraphics[width=0.31\textwidth]{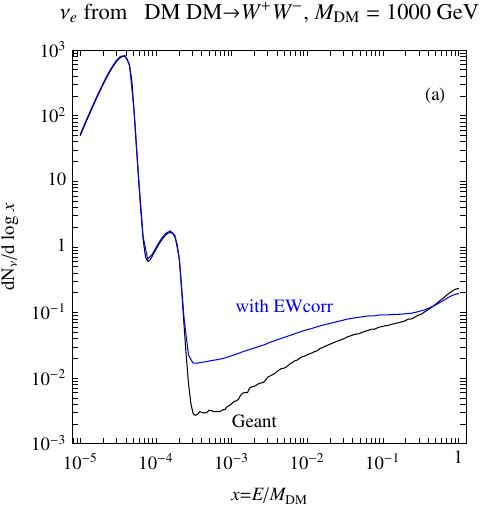} \quad
\includegraphics[width=0.31\textwidth]{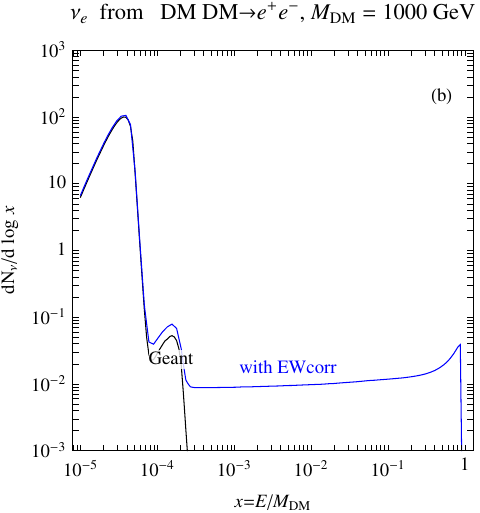} \quad
\includegraphics[width=0.31\textwidth]{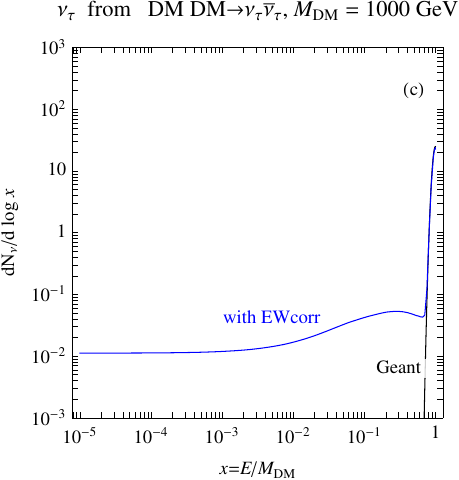}\\
\caption{\em Comparison of the neutrino spectra with and without the addition of {\bfseries ElectroWeak effects}, for a few representative cases.
\label{fig:EWcomparison}}
\end{center}
\end{figure}

In the previous subsection we have presented the results of the detailed MonteCarlo computation of neutrino spectra from DM DM annihilations into pairs of SM particles. There are however higher order effects that can be relevant and are not included in such computations.
In general, higher order corrections cannot be computed without having a full DM model: one needs to know the particles mediating the annihilation process in order to include all possible relevant diagrams. 
But {\em some} dominant higher order corrections {\em can} be computed in a model-independent way:
those enhanced by logarithms of ratios of particle masses, which describe bremsstrahlung.
In terms of Feynman diagrams, such process corresponds to attaching soft or collinear particles to the
SM final state particles. 
While the bremsstrahlung emission due to electromagnetic and strong interactions is automatically performed by MonteCarlo codes,
the bremsstrahlung emission due to electroweak interactions  is not included and is equally significant,
if $M_{\rm DM}\gtrsim M_W$.

We therefore include electroweak bremsstrahlung at leading order in the electroweak couplings by `post-processing' the output of the MonteCarlo as described in~\cite{EWDM}.
At large DM masses, such bremsstrahlung corrections are enhanced by $\ln (M_{\rm DM}/M_W)$ logarithms, which become large for $M_{\rm DM}\gg M_W$.
Such enhanced terms are model-independent: in our code we turn them on abruptly when $M_{\rm DM}\circa{>} M_W$.
In a full DM model,  these effects would actually appear in a smooth model-dependent way when increasing the DM mass. We instead neglect the finite non-logarithmic terms, that cannot be computed in a model-independent way.

\medskip

In practice, we proceed as follows.
In the previous section we computed  $dN^{\rm MC}_{J\to \nu_\ell}/dx$: the
MonteCarlo spectra in $x=E/M_{\rm DM}$ of $\nu_\ell$ produced by DM annihilations into a generic
two-body state $J$.
To include EW bremsstrahlung, we convolute such spectra with a set of electroweak splitting functions $D^{\rm EW}_{I\to J}(z)$ 
(probability that $I$ radiates a particle $J$ with energy reduced by a factor $z$) as follows:
\beq \label{eq:master}
\frac{dN_{I\to \nu_\ell}}{d\ln x} (M_{\rm DM},x)=\sum_J \int_x^1 dz \,D^{\rm EW}_{I\to J}(z)\;\frac{dN^{\rm MC}_{J\to \nu_\ell}}{d\ln x}\left(zM_{\rm DM},\frac{x}{z}\right),\eeq
The sum is over all EW splittings.
The rationale behind this procedure is that
splittings are kinematically different from decays and happen before decays, when particles have a large virtuality.
The splitting functions are predicted by the SM and listed in~\cite{EWDM}.
For example, left-handed electrons can radiate neutrinos via the EW splitting $e_L\to\nu_e W_T$ described by the function
\beq D^{\rm EW}_{e_L\to \nu_e}(z)=  \delta(1-z) \left[1 +  \frac{\alpha_2}{4\pi}  \bigg(\frac{3\ell}{2}-\frac{\ell^2}{2}\bigg)
\right] +
\frac{\alpha_2}{4\pi} \bigg( \frac{1+z^2}{1-z} L(1-z)\bigg)
. \eeq
where $\ell=\ln 4M_{\rm DM}^2/M_Z^2$ and
\begin{equation}\label{Lexact}
L(x)=
\ln\frac{M_{\rm DM}^2x^2}{M_Z^2}+2\,\ln\left(
1+\sqrt{1-\frac{M_Z^2}{M_{\rm DM}^2x^2}}
\right).
\end{equation}

\medskip 

From a phenomenological point of view, such effects can be relevant. Annihilation channels that would produce a low yield of neutrinos (such as DM DM $\to e^+ e^-$ or  DM DM $\to \mu^+ \mu^-$ where muons are severely slowed down or absorbed by matter) are significantly affected by EW bremsstrahlung because they receive contributions from channels with large yield of neutrinos (such as DM DM $\to W^+W^-$). This case  is shown in panel (b) of fig.\fig{EWcomparison}. The other panels in fig.\fig{EWcomparison} show other cases in which the impact of EW effects is sizable. For a channel like $W^+W^-$ (panel (a)), the radiation leads to a neutrino flux enhanced by about an order of magnitude at low energy. An `extreme' case is reproduced in panel (c): for annihilations directly into tau neutrinos and looking at the flux of $\nu_\tau$ themselves, EW corrections add to the na\"ive monochromatic spectrum a broad low energy shoulder, as a consequence of the decay of the EW gauge bosons emitted by the primary $\nu_\tau\bar\nu_\tau$ pair.

\subsection{Spectra at production: results}

\begin{figure}[p]
\parbox[b]{.31\linewidth}{
\includegraphics[width=\linewidth]{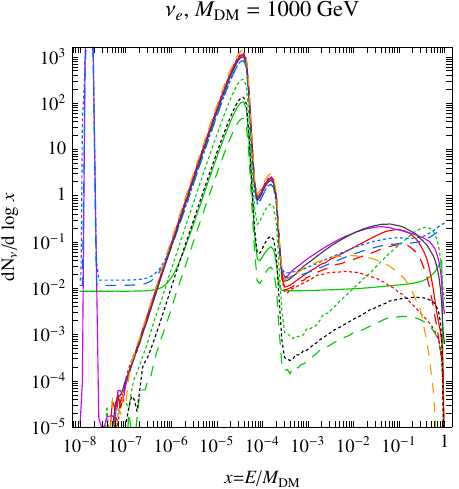}} \hfill
\parbox[b]{.31\linewidth}{
\includegraphics[width=\linewidth]{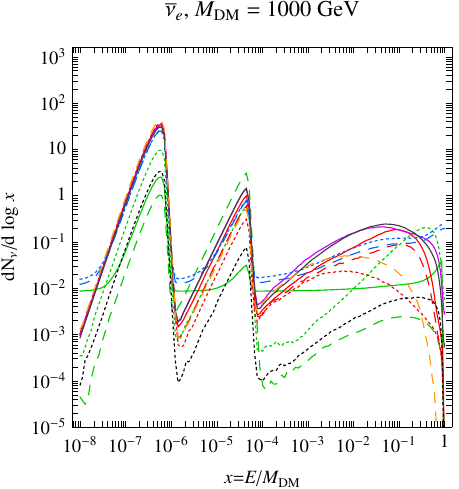}} \hfill
\parbox[b]{.31\linewidth}{
\includegraphics[width=\linewidth]{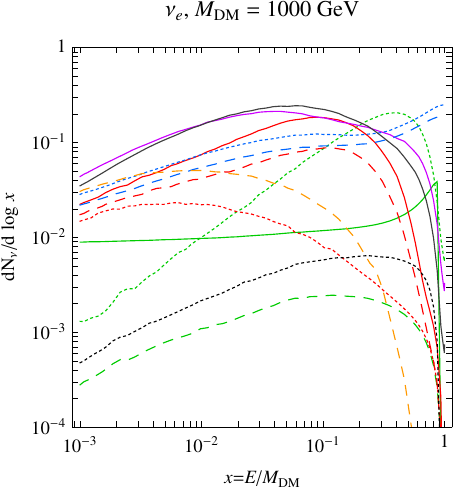}}\\[5mm]
\parbox[b]{.31\linewidth}{
\includegraphics[width=\linewidth]{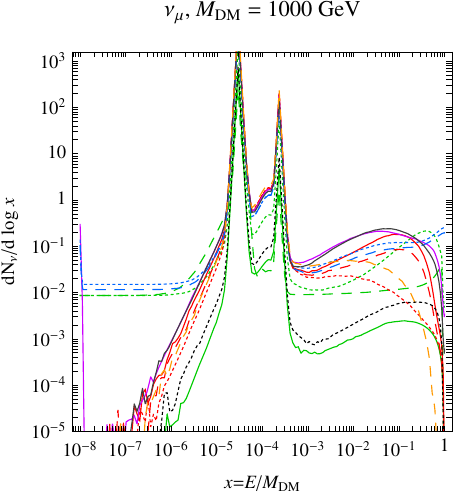}} \hfill
\parbox[b]{.31\linewidth}{
\includegraphics[width=\linewidth]{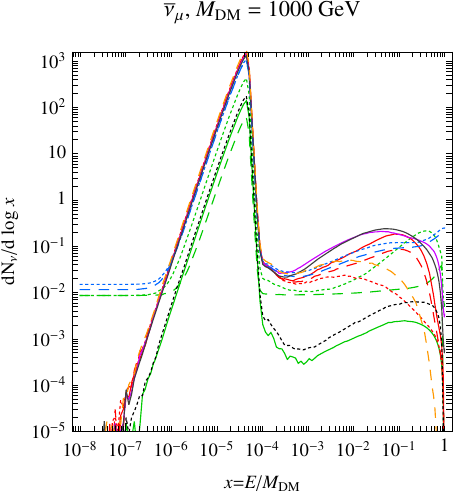}} \hfill
\parbox[b]{.31\linewidth}{
\includegraphics[width=\linewidth]{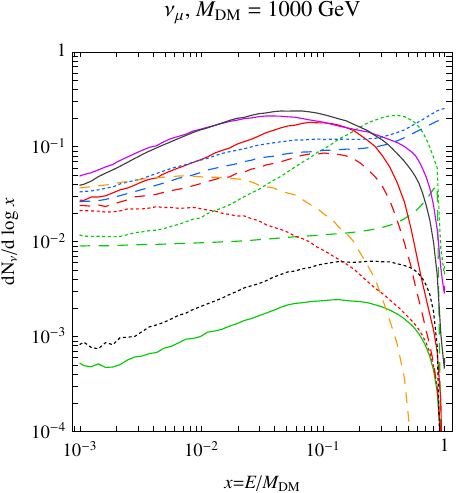}}\\[5mm]
\parbox[b]{.31\linewidth}{
\includegraphics[width=\linewidth]{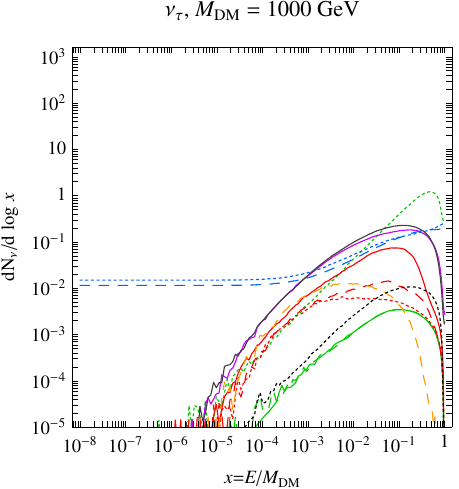}} \hfill
\parbox[b]{.31\linewidth}{
\includegraphics[width=\linewidth]{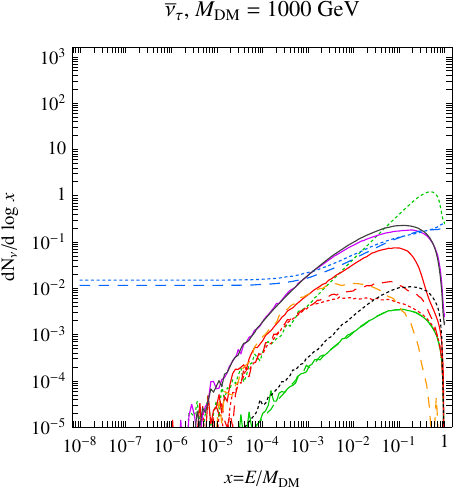}} \hfill
\parbox[b]{.31\linewidth}{
\includegraphics[width=\linewidth]{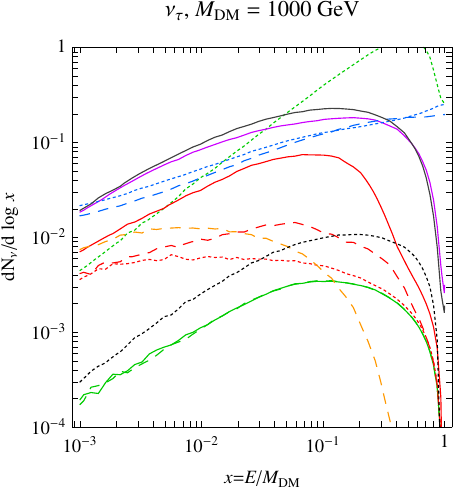}}\\[4mm]
\parbox[c]{.66\linewidth}{
\caption{\em 
\label{fig:finalspectraproduction} Final results for the {\bfseries neutrino spectra at production}, including all effects (in particular ElectroWeak corrections). Left column: neutrino spectra. Central column: antineutrino spectra. Right column: zoom on the high energy portion of the neutrino spectra. Upper row: $e$ flavor; middle row: $\mu$ flavor; bottom row: $\tau$ flavor. }} \hfill
\parbox[c]{.31\linewidth}{
\includegraphics[width=\linewidth]{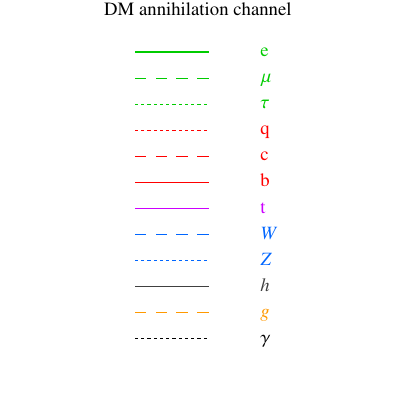}}
\end{figure}

Fig.~\ref{fig:finalspectraproduction} presents, for reference, an example of our final results for the neutrino spectra at production. We plot the spectra of $\nu$ (first column) and $\bar\nu$ (second column) for all the channels that we consider for a sample DM mass of 1 TeV. The spectra are normalized per one annihilation of two DM particles. 
The considered range of $x=E/M_{\rm DM}$ covers from $10^{-8}$ to $ 1$. 
In the third column we zoom on the high energy part, relevant for neutrino detectors such as {\sc IceCube}.

\medskip

All these spectra are provided in numerical form.


\section{Neutrino propagation}
\label{propagation}

Propagation of neutrinos from the interior of the Sun to the Earth is affected by flavor oscillations and (at energies above tens of GeV)
by Neutral Current (NC) and Charged Current (CC) interactions with solar matter, 
which give rise to absorption and (when a $\tau$ lepton is produced) to  regeneration of neutrinos with lower energies. 
In the following we discuss each of these effects and summarize the formalism we employ.

\subsection{Formalism}

Coherent flavor oscillations and  coherence-breaking interactions with matter  simultaneously affect neutrino propagation.
Following~\cite{earlyspectra}, the appropriate formalism that marries in a quantum-mechanically consistent way these two aspects, consists in studying the spatial evolution of the $3\times 3$ matrix of densities of neutrinos, ${\mb{\rho}}(E_\nu)$, and of anti-neutrinos, $\bar{\mb{\rho}}(E_\nu)$. The diagonal entries of the density matrix represent the population of the corresponding flavors, whereas the off-diagonal entries quantify the quantum superposition of flavors.\footnote{Alternatively, the fully numerical approach pursued in {\sc Wimpsim}~\cite{WIMPSIM} consist in writing down an event-based MonteCarlo that follows the path of a single neutrino undergoing oscillations and interactions (with given probabilities). The two approaches yield results which are very well in agreement, for any practical purpose.}  
The matrices $\rhob(E_\nu)$ and $\bar\rhob(E_\nu)$
satisfy a coupled system of integro-differential equations in the distance $r$ from the center of the Sun:
\beq \label{eq:drho}\frac{d\rhob}{dr} =
- i [\mb{H},\ \rhob] +
\left.\frac{d\rhob}{dr}\right|_{\rm NC}+
\left.\frac{d\rhob}{dr}\right|_{\rm CC}
\eeq
with an analogous equation for $\bar\rhob$.
\begin{itemize}
\item 
The first term describes oscillations, computed including the vacuum mixing and the MSW matter effect~\cite{MSW}.
The effective Hamiltonian reads
\beq\label{eq:H}
 \mb{H} =\frac{\mb{m}^\dagger \mb{m}}{2E_\nu } \pm
\sqrt{2} G_{\rm F}\bigg[N_e\ \diag\mb{(}1,0,0\mb{)} -\frac{N_n}{2}\ \diag\mb{(}1,1,1\mb{)}\bigg]\ ,
\eeq
where $\mb{m}$ is the $3 \times 3$ neutrino mass matrix, and the $+$ ($-$) sign
applies for neutrinos (anti-neutrinos).
One has $\mb{m}^\dagger \mb{m} = \mb{V}\cdot{\rm diag}(m_1^2,m_2^2,m_3^2)\cdot \mb{V}^\dagger$
where $m_{1,2,3}>0$ are the neutrino masses and $\mb{V}$ is the neutrino mixing matrix
given by \begin{equation} \mb{V} =
R_{23}(\theta_{23}) \cdot
R_{13}(\theta_{13}) \cdot
\hbox{diag}\,(1,  e^{i \phi},1) \cdot
R_{12}(\theta_{12}) 
\label{eq:Vunitary}
\end{equation}
where $R_{ij}(\theta_{ij})$ represents a
rotation by $\theta_{ij}$ in the $ij$ plane and
we assume the present best fit values for the mixing parameters~\cite{nuosc}\footnote{We neglect the indications possibly in favor of a non-maximal $\theta_{\rm atm}$ and we do not consider the small dependence of the best fit values on the choice of the mass hierarchy.}
$$\tan^2\theta_{\rm sun}=0.45,\qquad 
\theta_{\rm atm}=45^\circ,\qquad \theta_{13}=8.8^\circ,$$
$$\Delta m^2_{\rm sun}={7.5}~10^{-5}\eV^2,\qquad
|\Delta m^2_{\rm atm}|={2.45}~10^{-3}\eV^2.$$

\item
The second  term in eq.\eq{drho} describes the absorption and re-emission 
due to NC scatterings $\nubarnu N \leftrightarrow \nubarnu N^*$  (where $N$ is any nucleon in the Sun and with $N^*$ we denote its possible excited state after the collision), which
remove a neutrino from the flux and re-inject it with a lower energy. 
So they contribute to the evolution equation as:
\beq\label{eq:NC}\left.\frac{d\rhob}{dr}\right|_{\rm NC} = - \int_0^{E_\nu} dE'_\nu 
\frac{d\Gamma_{\rm NC}}{dE'_\nu} (E_\nu,E'_\nu) \rhob(E_\nu)+
\int_{E_\nu}^\infty dE'_\nu 
\frac{d\Gamma_{\rm NC}}{dE_\nu} (E'_\nu,E_\nu) \rhob(E'_\nu)\eeq
where
\beq\label{eq:Gamma}
\Gamma_{\rm NC}(E_\nu,E'_\nu) = N_p(r)\ \sigma(\nu_\ell p\to \nu_\ell' X)
+N_n(r)\ \sigma(\nu_\ell n\to \nu_\ell' X).
\eeq

\item The third term in eq.\eq{drho} describes Charged Current (CC) scatterings $\nubarnu_\ell N \rightarrow \ell^\pm X$ of an initial neutrino $\nu_\ell$ with energy $E_{\nu}$, which
remove the $\nu_\ell$ from the flux and produce a charged lepton $\ell$ and scattered hadrons $X$.
They decay back into neutrinos $\nu_{\ell'}$ and anti-neutrinos $\bar\nu_{\ell'}$ with lower energy $E'_\nu$:
their energy distributions are described by the function $f_{\ell\to\ell'}(E_\nu, E'_\nu)$.
When the initial neutrino is $\nu_\tau$ ($\bar\nu_\tau$), the produced $\tau^-$ ($\tau^+$) decays promptly before losing energy,
giving rise to energetic $\nu_\tau,\bar\nu_e,\bar\nu_\mu$ ($\bar\nu_\tau,\nu_e,\nu_\mu$): this is the tau regeneration phenomenon~\cite{taureg}.
When instead the initial neutrino is a $\nu_e$ or $\nu_\mu$ we assume that the produced $e, \mu$ is totally absorbed  and we neglect the corresponding low energy neutrinos.
CC scatterings thereby affect the propagation of neutrinos with the term
\begin{eqnarray}
\left.\frac{d\rhob}{dr}\right|_{\rm CC} &=& - \frac{\{\mb{\Gamma}_{\rm CC},\rhob\}}{2}+
\int \frac{dE^{\rm in}_\nu}{E^{\rm in}_\nu}  
 \bigg[ \mb{\Pi}_\tau \rho_{\tau\tau}(E^{\rm in}_\nu) \Gamma_{\rm CC}^\tau(E^{\rm in}_\nu) 
 f_{\tau\to\tau}({E_\nu^{\rm in}},{E_\nu}) \nonumber \\
 & &\qquad
 + \mb{\Pi}_{e,\mu} \bar\rho_{\tau\tau} (E^{\rm in}_\nu) \bar\Gamma_{\rm CC}^{\tau} (E^{\rm in}_\nu)
 f_{\bar\tau\to e,\mu}( E^{\rm in}_\nu, E_\nu)
\bigg],  
\end{eqnarray}
where $\mb{\Pi}_\ell$ is the projector on the flavor $\nu_\ell$: e.g.\ $\mb{\Pi}_e = \diag\mb{(}1,0,0\mb{)}$.
The matrices $\mb{\Gamma}_{\rm CC}$, $\bar{\mb{\Gamma}}_{\rm CC}$  that describe the rates of CC interactions are given by
$\mb{\Gamma}_{\rm CC}(E_\nu) = \diag\mb{(}\Gamma_{\rm CC}^e,
\Gamma_{\rm CC}^\mu,\Gamma_{\rm CC}^\tau\mb{)}$, where
\beq
\Gamma_{\rm CC}^\ell=
N_p(r)\  \sigma(\nu_\ell p\to \ell X)
+N_n(r)\  \sigma(\nu_\ell n\to \ell X).
\eeq
\end{itemize}

\noindent In both NC and CC processes, we neglect low energy neutrinos that might emerge from the scattered hadrons and light leptons, i.e. the de-excitation of $N^*$ in NC and the decay of $X$ and $e, \mu$ in CC. E.g. in particular we neglect the very low energy neutrinos with $E_\nu \sim m_{\pi, K, \mu}$ coming from the decay at rest of light hadrons/leptons. In order to include them, one should implement neutrino/matter interactions in dedicated codes such as {\sc Geant}, which currently do not include them. I.e. this would be analogous to the work we performed in \ref{Geant}, now with neutrinos as primary particles. We postpone this to possible future improvements. We estimate that such a neglected effect would only give a small enhancement in the final flux of neutrinos at very low energy.   

\subsection{Transition probabilities}

\begin{figure}
\begin{center}
$$\includegraphics[width=0.95\textwidth]{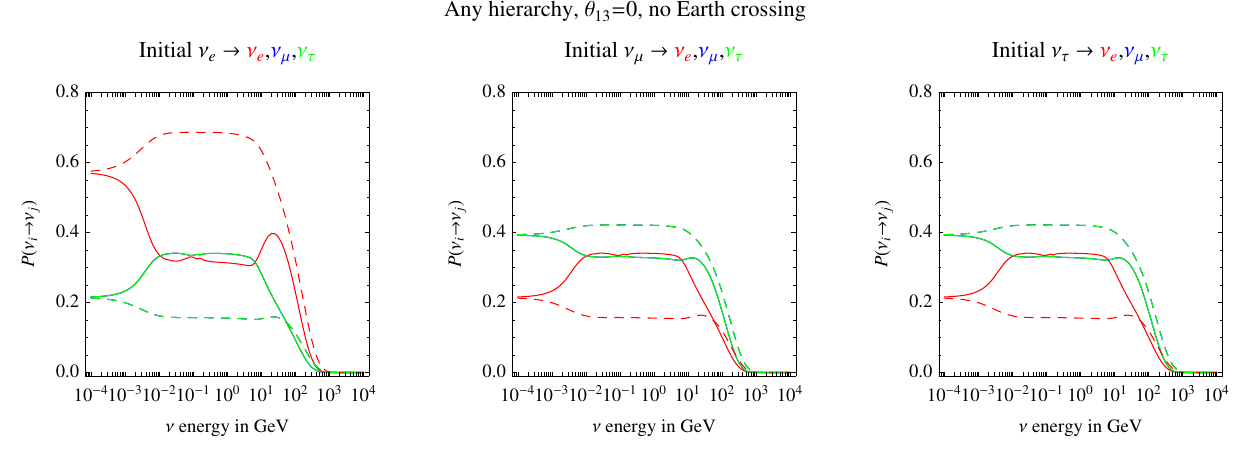}$$
\vspace{-1cm}
\caption{\em Neutrino (continuous curves) and anti-neutrino (dashed) {\bfseries transition probability from the Sun to the Earth}, assuming $\theta_{13}=0$. At $E\gg 10\GeV$ the total probability $P(\nu_i \to \sum_f \nu_f)$ is smaller than 1  because of absorption. 
Since $theta_{13} = 0$ in this figure, $\nu_\mu$ and $\nu_\tau$ are maximally mixed and their lines overlap. The probabilities plotted here do not include regenerated neutrinos, see text for details. 
\label{fig:nuprop0}}
\vspace{0.4cm}
$$\includegraphics[width=0.95\textwidth]{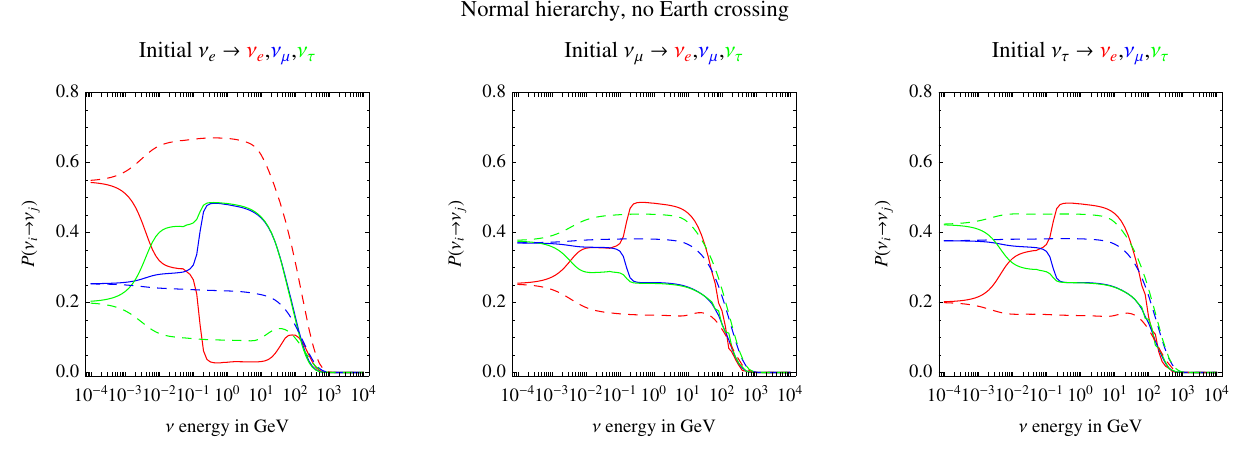}$$
$$\includegraphics[width=0.94\textwidth]{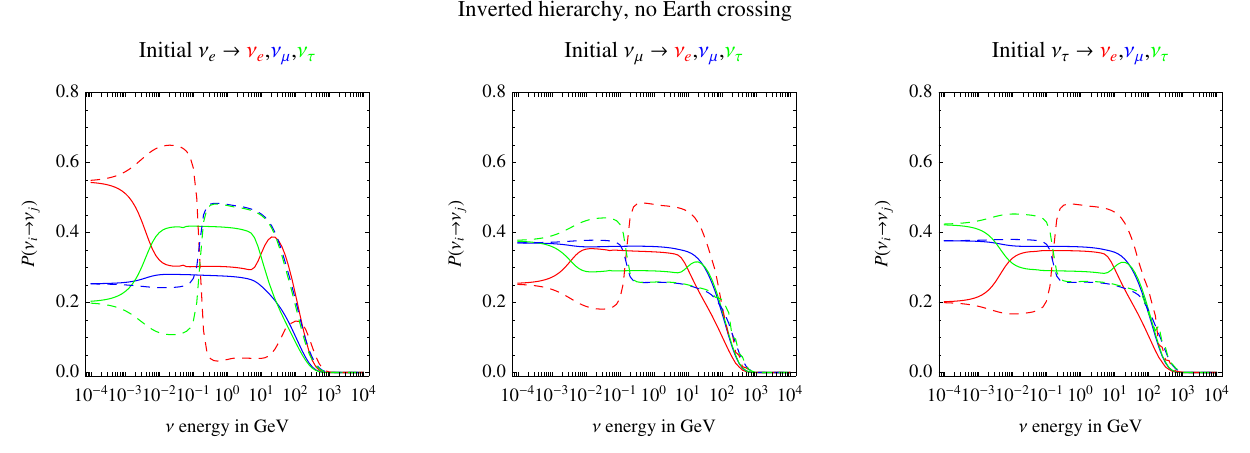}$$
\vspace{-1cm}
\caption{\em Like fig.\fig{nuprop0}, but allowing for the actual, non-vanishing value of $\theta_{13}$ and therefore distinguishing the neutrino mass hierarchies.
\label{fig:nuprop}}
\end{center}
\end{figure}

We numerically solve the full evolution equation (eq.~\ref{eq:drho}) starting from the initial condition dictated
by the spatial distribution of DM annihilations inside the Sun,
proportional to $n(r)^2$ given by eq.\eq{n(r)}. In practice, we numerically computed the full transition probabilities $P_\pm(\nu_\ell(E')\to \nu_i (E))$ from the Sun to the Earth,
with $\ell= \{e,\mu,\tau,\bar e,\bar \mu,\bar\tau\}$, $i = \{1,2,3,\bar 1,\bar 2,\bar 3\}$ and $E\le E'$
in the two cases of normal $(P_+$) and of inverted ($P_-$) neutrino mass hierarchy.

\medskip

The transition probabilities incorporate all propagation effects. In order to have a qualitative understanding of their behavior, however,
we plot them only in the limit $E = E'$ that excludes the neutrinos produced by regeneration (at $E < E'$).

Fig.\fig{nuprop0} and\fig{nuprop} show the transition probabilities $P(\nu_\ell \to \nu'_\ell)$ between flavour eigenstates
of neutrinos (continuous curves) and anti-neutrinos (dashed) from the Sun to the surface of the Earth (before a possible crossing of the Earth, which will be considered in the next subsection).
Fig.\fig{nuprop0} shows the probabilities computed for $\theta_{13}=0$.
In fig.\fig{nuprop}, $\theta_{13}$ is restored to its actual value $\theta_{13}= 8.8^\circ$ and therefore we need to distinguish the hierarchy: the upper row refers to a normal spectrum of neutrinos ($m_1\ll m_2\ll m_3$), while the lower row to an inverted spectrum
($m_1 \ll m_2 \approx m_3$).

Considering for example $P(\nu_e\to\nu_e)$: it goes from $1- \frac{1}{2} \sin^22\theta_{12}$ at $E\ll\MeV$ (averaged vacuum oscillations)
to $\sin^2\theta_{12}$ at larger energies (adiabatic MSW resonance for $\theta_{12}$).
If neutrinos have normal hierarchy, at $E_\nu\gg 100\MeV$ also $\theta_{13}$ is enhanced by an adiabatic MSW resonance, 
and $P(\nu_e\to\nu_e)$ drops.\footnote{The MSW resonance occurs in a region with width in matter density given by
$\Delta n \approx  2\theta n$, which is smaller for smaller $\theta$.
Numerical solutions to eq.\eq{drho} obtained using the Runge-Kutta method
must thereby employ a smaller step in $r$ when $\theta_{13}$ is turned on.}
Many effects happen when $E_\nu \sim 10\GeV$: MSW resonances cease to be adiabatic, and the solar oscillation wave-length becomes
comparable to the size of the Sun.  These two effect cause an increase of $P(\nu_e\to\nu_e)$ towards its vacuum oscillation value.
However this increase gets stopped by neutrino absorption due to interactions with solar matter, which causes all probabilities to drop to zero
in the limit of large energy.
The other oscillation probabilities can be similarly understood.  

Our propagation results agree, in the limit cases, with known results of the past, and in particular with~\cite{nuDM1} and the works that made use of it.  
We verified that the unknown neutrino CP-violating phase that enters into oscillations has negligible effects.
Furthermore, such results do not depend on the Majorana or Dirac nature of neutrinos.

\subsection{Earth crossing}

The spectra of neutrinos and anti-neutrinos at the Earth are finally computed (in terms of mass eigenstates $\nu_i$) by convoluting the transition probabilities with the spectra at production as
\beq \frac{dN_{\nu_i}^\pm}{dE} = \sum_\ell \int_{E}^M dE' ~P_\pm(\nu_\ell(E')\to \nu_i (E)) \frac{dN_{\nu_\ell}^{\rm prod}}{dE'}.\eeq

The final step consists in taking into account the oscillations in the matter of the Earth. If neutrinos do not cross the Earth, the energy spectra for the neutrino flavour eigenstates are simply given by
\beq \frac{dN^\pm_{\nu_\ell}}{dE_\nu } = \sum_i |V_{\ell i}|^2  \frac{dN^\pm_{\nu_i}}{dE_\nu }.\eeq
If instead neutrinos cross the Earth with zenith angle $\vartheta$ ($\cos\vartheta=-1$ corresponds to the maximal vertical crossing,
and $\cos\vartheta=0$ corresponds to the minimal horizontal crossing), the neutrino fluxes at detection are given by
\beq \frac{dN^\pm_{\nu_\ell}}{dE_\nu } = \sum_i  P^\pm_{\rm earth}(\nu_i \to \nu_\ell, E_\nu,\vartheta) \frac{dN^\pm_{\nu_i}}{dE_\nu }.\eeq
where the oscillation probabilities $P_{\rm earth}$ are readily computed adopting the standard Earth density model. We neglect neutrino absorption within the Earth  (they would be relevant only for energies above $\sim$ 10 TeV and neutrinos with those energy essentially do not emerge from the Sun, as discussed above).

\begin{figure}
\begin{center}
$$\includegraphics[width=\textwidth]{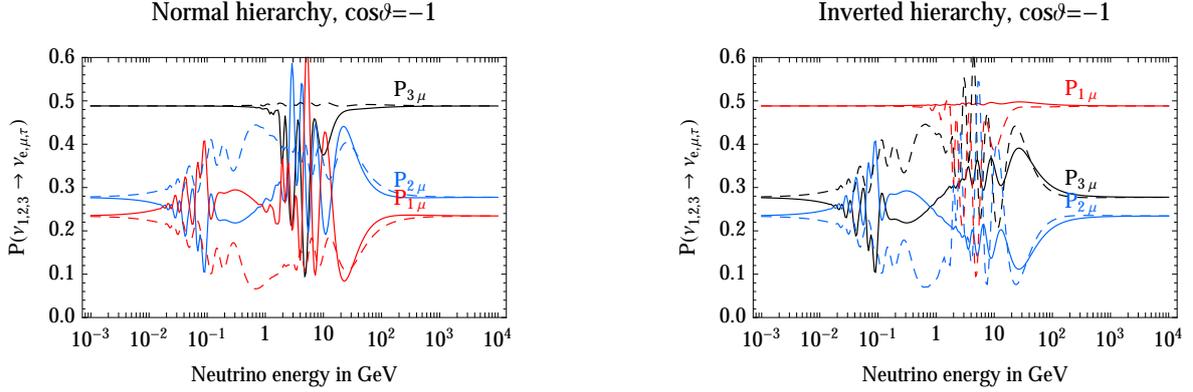}$$
\caption{\em {\bfseries Earth crossing oscillation probabilities} into $\nu_\mu$ (continuous curves) and into $\bar\nu_\mu$ (dashed), for neutrinos crossing vertically the Earth.
\label{fig:Pearth}}
\end{center}
\end{figure}

Some Earth oscillation probabilities are plotted in fig.\fig{Pearth} for illustration; for large and small neutrino energy they approach the limiting values
$|V_{\ell i}|^2$.

\section{Final result}
\label{sec:finalresult}

\begin{figure}[p]
\parbox[b]{.31\linewidth}{
\includegraphics[width=\linewidth]{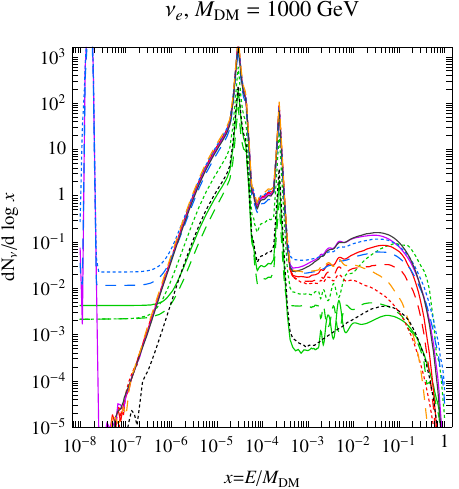}} \hfill
\parbox[b]{.31\linewidth}{
\includegraphics[width=\linewidth]{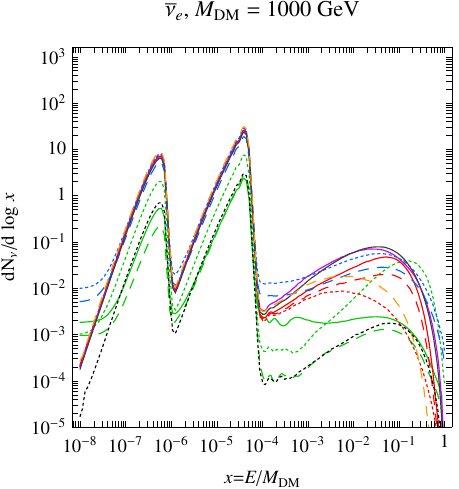}} \hfill
\parbox[b]{.31\linewidth}{
\includegraphics[width=\linewidth]{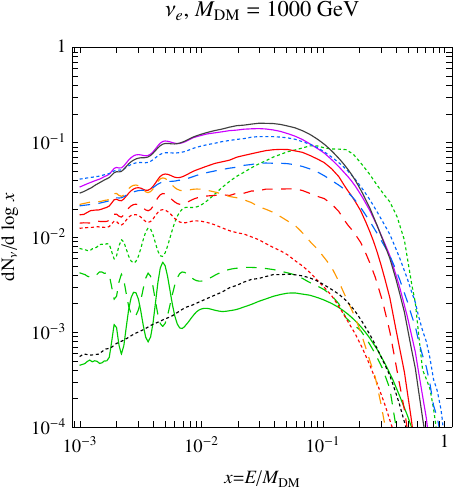}}\\[5mm]
\parbox[b]{.31\linewidth}{
\includegraphics[width=\linewidth]{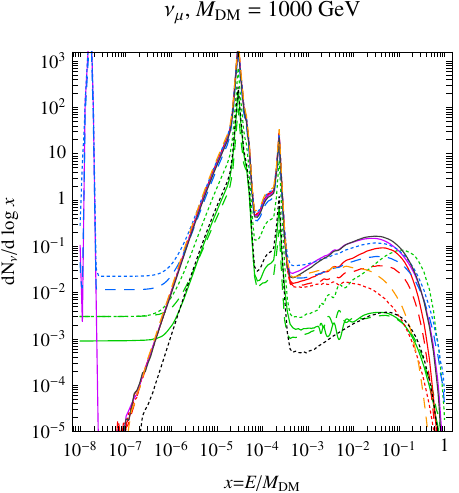}} \hfill
\parbox[b]{.31\linewidth}{
\includegraphics[width=\linewidth]{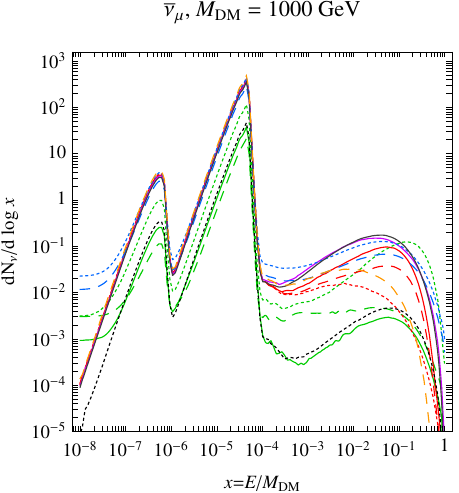}} \hfill
\parbox[b]{.31\linewidth}{
\includegraphics[width=\linewidth]{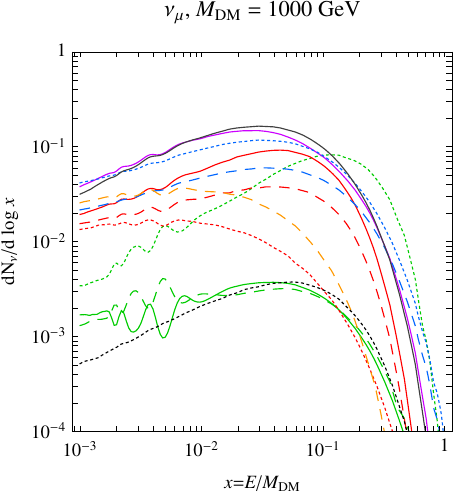}}\\[5mm]
\parbox[b]{.31\linewidth}{
\includegraphics[width=\linewidth]{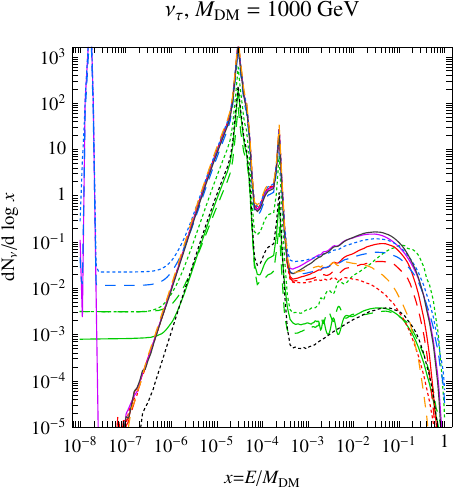}} \hfill
\parbox[b]{.31\linewidth}{
\includegraphics[width=\linewidth]{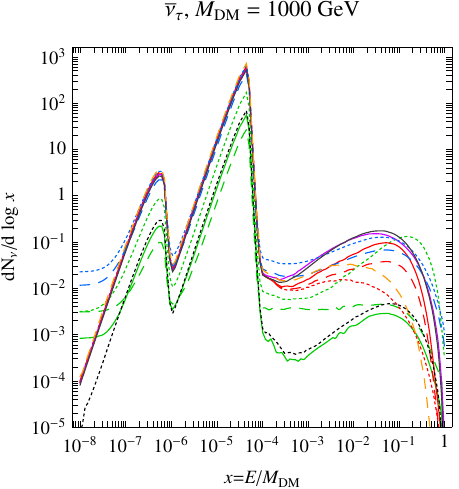}} \hfill
\parbox[b]{.31\linewidth}{
\includegraphics[width=\linewidth]{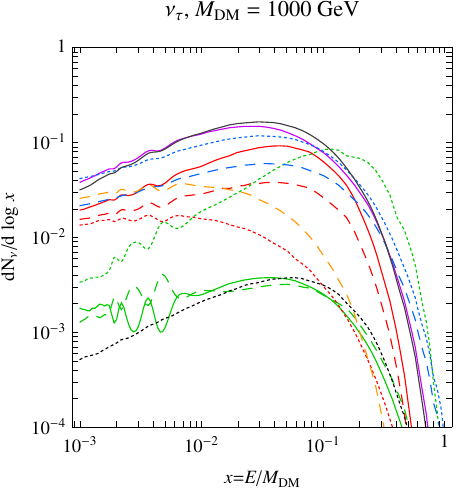}}\\[4mm]
\parbox[c]{.66\linewidth}{
\caption{\em 
\label{fig:finalspectradetection} Final results for the {\bfseries neutrino spectra at detection}, including all propagation effects. For definiteness we choose the case of Normal Hierarchy and neutrinos crossing vertically the Earth. Left column: neutrino spectra. Central column: antineutrino spectra. Right column: zoom on the high energy portion of the neutrino spectra. Upper row: $e$ flavor; middle row: $\mu$ flavor; bottom row: $\tau$ flavor. These plots can be directly compared with those in fig.\fig{finalspectraproduction}.}} \hfill
\parbox[c]{.31\linewidth}{
\includegraphics[width=\linewidth]{figs/legenda.pdf}}
\end{figure}

\begin{figure}[p]
\parbox[b]{.31\linewidth}{
\includegraphics[width=\linewidth]{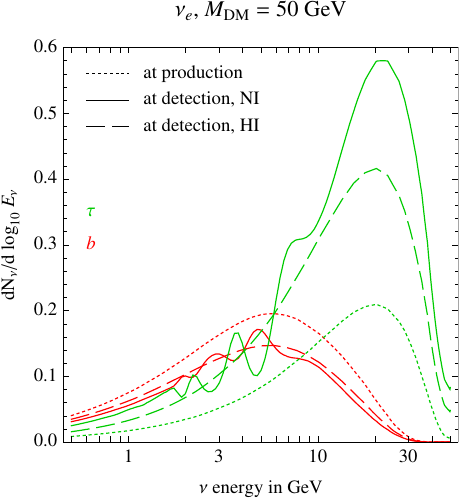}} \hfill
\parbox[b]{.31\linewidth}{
\includegraphics[width=\linewidth]{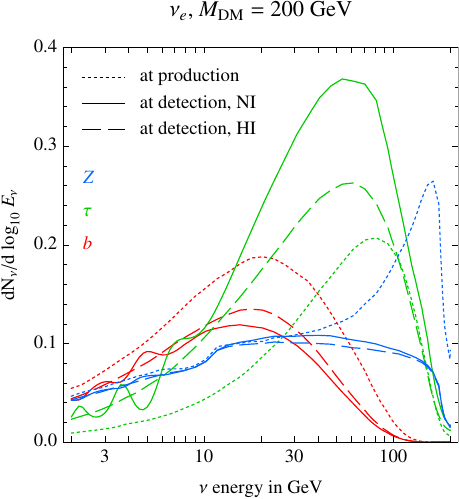}} \hfill
\parbox[b]{.31\linewidth}{
\includegraphics[width=\linewidth]{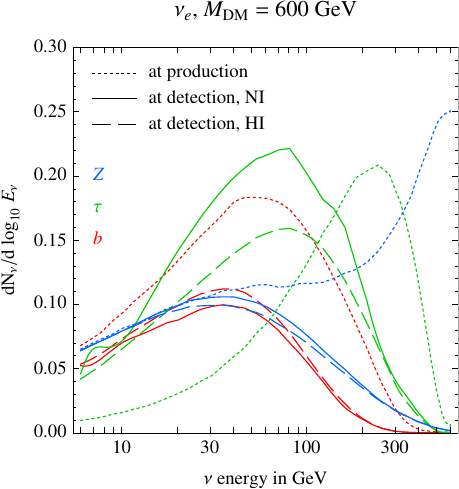}}\\[5mm]
\parbox[b]{.31\linewidth}{
\includegraphics[width=\linewidth]{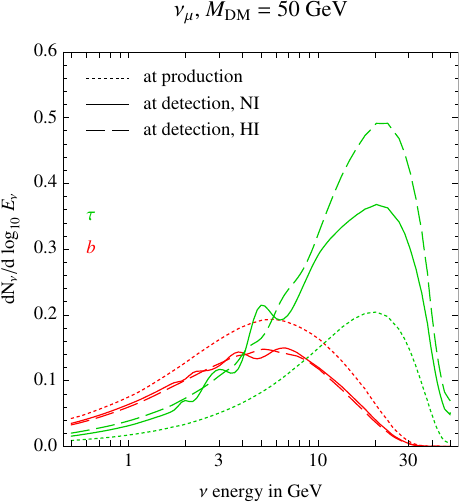}} \hfill
\parbox[b]{.31\linewidth}{
\includegraphics[width=\linewidth]{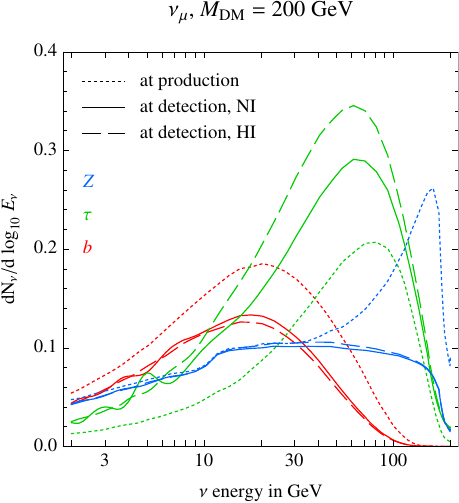}} \hfill
\parbox[b]{.31\linewidth}{
\includegraphics[width=\linewidth]{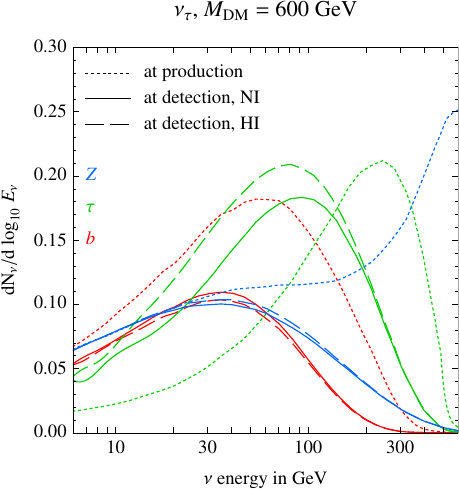}}\\[5mm]
\parbox[b]{.31\linewidth}{
\includegraphics[width=\linewidth]{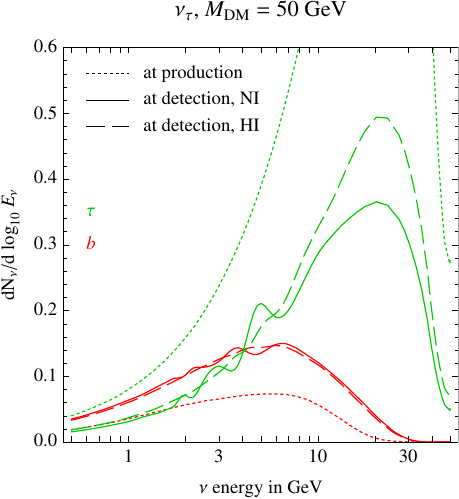}} \hfill
\parbox[b]{.31\linewidth}{
\includegraphics[width=\linewidth]{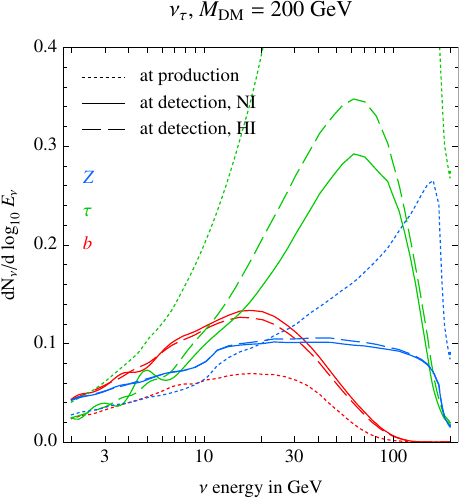}} \hfill
\parbox[b]{.31\linewidth}{
\includegraphics[width=\linewidth]{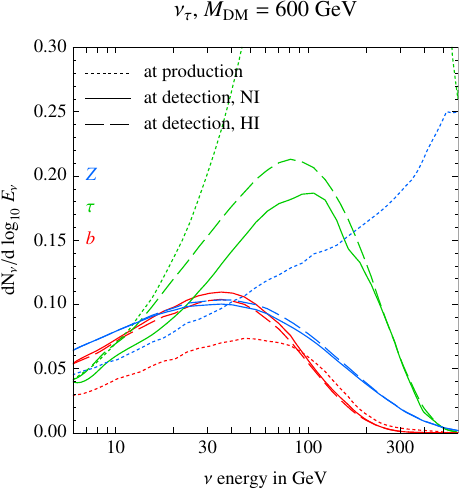}}\\[4mm]
\caption{\em \label{fig:finalfluxescompared}  {\bfseries Comparison of the neutrino spectra} at production and detection,  showing the effects of propagation. For definiteness we choose the case of neutrinos crossing vertically the Earth. Upper row: $e$ flavor; middle row: $\mu$ flavor; bottom row: $\tau$ flavor. Different columns: different values of the DM mass.}
\end{figure}

Recapping the results of the previous sections: we provide the energy spectra at detection of neutrino flavor eigenstates  $dN^\pm_{\nu_\ell}/dE_\nu$ produced by one DM annihilation in the Sun, after taking into account all effects that neutrinos experience during their journey.
We provide two sets of spectra: $dN^+_{\nu_i}/dE_\nu$ corresponding to neutrinos with normal mass hierarchy,
and $dN^-_{\nu_i}/dE_\nu$ corresponding to neutrinos with inverted mass hierarchy.

Fig.\fig{finalspectradetection} presents, for reference, an example of our final results for the neutrino spectra at detection, analogously to fig.\fig{finalspectraproduction}. The spectra are, as always, normalized per one annihilation of two DM particles. 

In fig.\fig{finalfluxescompared} we present a more detailed comparison of the effect of propagation, for a few selected masses and channels. One sees, for instance:
\begin{itemize}
\item The effect of flavor vacuum oscillations: for an annihilation into $\tau^+\tau^-$ the flux of electron and muon neutrinos is greatly enhanced and the corresponding flux of tau neutrinos is depleted; for an annihilation into $\bar b b$, the opposite happens since $\nu_{e,\mu}$ mostly emerge from the $b$ channel.
\item The effect of solar matter absorption: moving towards higher masses, the spectra are significantly degraded in energy; the case of the $Z$ spectrum (peaked at production) is the most apparent. For even larger $m_{\rm DM}$ all spectra approach a limit, `bell-shaped' exponential spectrum dictated by the maximum energy to which the Sun is transparent~\cite{nuDM1}.
\item The effect of Earth crossing oscillations: the wiggles at around 1 to 10 GeV.
\end{itemize}

\noindent All these spectra are provided in numerical form.

\medskip

The neutrino fluxes can then be converted into fluxes of detectable particles (up-going muons, through-going muons, showers). 
Various experiments searched for DM neutrinos from the Sun, producing bounds:
{\sc SuperKamiokande}~\cite{Tanaka:2011uf},
{\sc Icecube}~\cite{Aartsen:2012kia},
{\sc Antares}~\cite{Adrian-Martinez:2013ayv} and
{\sc Baksan}~\cite{Boliev:2013ai}.
These collaborations report results as bounds on DM annihilations assuming some DM annihilation channel
and rates of events (such as through-going-muons) with cuts optimised assuming specific energy spectra.
Unfortunately, however,  they do not presently report experimental bounds on the {\em neutrino} fluxes from the Sun
(this kind of analysis could be done assuming monochromatic neutrino spectra at different energies).\footnote{A step in this direction has been attempted in~\cite{Scott:2012mq}.} 
Therefore we cannot compare our improved neutrino fluxes with experimental data
and we cannot at present derive improved bounds on all the DM annihilation channels.

\section{Conclusions}
\label{sec:concl}

In this paper we have reconsidered the computation of the high energy neutrino fluxes from the annihilation of DM particles captured in the Sun. With respect to other DM indirect detection search strategies, such a signature is particularly interesting and timely given that: i) It is unique in the sense that it cannot be mimicked by known astrophysical processes; it is therefore virtually background-free at the source, except for solar corona neutrinos~\cite{corona} (the background for detection consists of atmospheric neutrinos). ii) The propagation of neutrinos from the Sun to the Earth is under control, in particular now that the neutrino oscillation parameters are (almost) all measured and with good precision. iii) The current and upcoming generation of neutrino telescopes are reaching the sensitivity necessary to probe one of the most interesting portions of the parameter space, in competition with other DM search strategies.

\bigskip

We have reviewed the capture of DM particles in the Sun, considering the relevant elements in solar matter and different assumptions for the halo velocity distribution. 

We have computed the neutrino spectra from the annihilation into all two-body SM annihilation channels, implementing a detailed description of the energy losses of primary particles in solar matter, including secondary neutrinos produced by the scattered matter and low energy ones from the decay at rest of light hadrons/leptons. We performed this computation with the {\sc Geant} MonteCarlo, after checking against {\sc Pythia}. We added on top of this the effect of ElectroWeak radiation, which is relevant for DM masses above some hundreds of GeV. We span the range of masses from 5 GeV to 100 TeV. 

We then perform the propagation of neutrinos from the center of the Sun to the detector,  taking into account (vacuum and matter) oscillations and interaction with solar matter, as well as Earth crossing. We adopt up-to-date oscillation parameters and consider normal or inverted neutrino mass hierarchy.

\bigskip

The main numerical outputs of the computation are given on the \myurl{www.marcocirelli.net/PPPC4DMID.html}{PPPC 4 DM ID website} (`DM$\nu$' section). We provide:
\begin{itemize}
\item[$\triangleright$] a numerical function for $\Gamma_{\rm ann}$,
\item[$\triangleright$] the neutrino spectra at production, including EW radiation effects,
\item[$\triangleright$] the neutrino spectra at detection, including all propagation effects.
\end{itemize}
These results can be readily used to derive predictions for the experimental observables. 

\bigskip

\small
\subsubsection*{Acknowledgments}
We thank Nicolao Fornengo for useful discussions and a patient cross-check of some of the results of Cirelli {\it et al.} in~\cite{nuDM1}. We thank Paolo Panci, Teresa Montaruli and Sofia Vallecorsa for useful discussions and Eugenio Del Nobile for pointing out a typo. We thank Vladimir Kulikovskiy for pointing out a typo in eq.~(\ref{eq:GammaCaptApproxNumbers}) of the v2 of this work, now corrected.\\
M.C. acknowledges the hospitality of the Institut d'Astrophysique de Paris ({\sc Iap}) and of the Theory Unit of {\sc Cern} where a part of this work was done.

\medskip

\footnotesize
\noindent Funding and research infrastructure acknowledgements: 
\begin{itemize}
\item[$\ast$] {\sc Esf} (Estonian Science Foundation) grant MTT8, grant 8499 and SF0690030s09 project,
\item[$\ast$] European Programme PITN-GA-2009-23792 ({\sc Unilhc}), 
\item[$\ast$] European Programme {\sc Erdf} project 3.2.0304.11-0313 Estonian Scientific Computing Infrastructure ({\sc Etais}),
\item[$\ast$] European Research Council ({\sc Erc}) under the EU Seventh Framework Programme (FP7/2007-2013)/{\sc Erc} Starting Grant (agreement n.\ 278234 --- `{\sc NewDark}' project) [work of M.C.],
\item[$\ast$] French national research agency {\sc Anr} under contract {\sc Anr} 2010 {\sc Blanc} 041301.
 \end{itemize}

\appendix
 
\end{document}